\documentclass[aps,pra,eqsecnum,superscriptaddress,groupedaddress]{revtex4}
\usepackage{amssymb,amsmath}
\usepackage{subdepth}
\usepackage{graphicx}
\usepackage{sansmath}
\usepackage{bm}
\usepackage{soul}
\usepackage{tikz}
\usepackage{upgreek}

\usepackage[breaklinks=true]{hyperref}
\hypersetup{
	colorlinks   = true, %Colours links instead of ugly boxes
	urlcolor     = blue, %Colour for external hyperlinks
	linkcolor    = red, %Colour of internal links
	citecolor   = green %Colour of citations
}

\DeclareMathOperator{\tr}{tr}

\newcommand{\ket}[1]{\vert#1\rangle}
\newcommand{\bra}[1]{\langle#1\vert}
\newcommand{\braket}[2]{\left\langle #1 | #2 \right\rangle}
\newcommand{\proj}[1]{| #1\rangle\!\langle #1 |}

\newcommand{\inv}{{\,\text{-}\hspace{-1pt}1}}
\newcommand{\ad}{\mathrm{ad}}

\newcommand{\Z}{\mathcal{Z}}

\renewcommand{\L}{\mathcal{L}}

\newcommand{\D}{\mathcal{D}}

\newcommand{\Linv}[1]{\underrightarrow{#1}}
\newcommand{\Rinv}[1]{\underleftarrow{#1}}

\newcommand{\R}{\mathbb R}
\newcommand{\C}{\mathbb C}
\newcommand{\U}{\mathrm U}
\newcommand{\SO}{\mathrm{SO}}

\newcommand{\SL}{\mathrm{SL}}

\newcommand{\g}{\mathfrak g}

\newcommand{\Prob}{\mathrm{Prob}}

\renewcommand{\O}{\mathcal{O}}
\newcommand{\Odot}{{\mathcal{O}\hspace{-0.575em}\raisebox{0.5pt}{$\boldsymbol{\cdot}$}\hspace{.275em}}}

\usepackage{color}

\begin{document}
	
	\title{Quantum Measurement and Continuous Markov Processes}
	
	\author{Christopher S. Jackson}
	\email{omgphysics@gmail.com}
	\affiliation{Perimeter Institute, Waterloo, Ontario N2L 6B9 Canada}

	\date{\today}
	
\begin{abstract}
These are the lecture notes for a course on diffusive quantum measuring instruments.
They were prepared and delivered at the Perimeter Institute on Mondays and Thursdays, from 2:30 to 4:00 PM, beginning October 27th, 2025 and ending December 11th, 2025.
These lectures were recorded and can be found at \textbf{https://pirsa.org/c25038}.
\end{abstract}
	
	%\pacs{}
	\maketitle

\tableofcontents

\subsection*{Five Examples of Diffusive Measurement}

\begin{enumerate}
	\item  von Neumann Measurement of a Single Hermitian Observable, \textsection \ref{indecisive}
	\item  Indirect GGW Heterodyne, \textsection \ref{Ex2}
	\item  Indirect GGW Homodyne, \textsection \ref{Ex3.1}, \textsection \ref{Ex3.2}, and \textsection \ref{Ex3.3}
	\item  Barchielli's Simultaneous P \& Q Measurement (SPQM), \textsection \ref{Ex4}
	\item  The Isotropic Spin Measurement (ISM), \textsection \ref{Ex5}
\end{enumerate}

\pagebreak 

\section*{Lecture Series Collection (https://pirsa.org/c25038)}

\begin{enumerate}
	\setcounter{enumi}{-1}
	\footnotesize 
	\item  Oct 21: ``The Planimeter and Contact Transformation: A Perfect Embodiment of the Weyl-Heisenberg Group and\\ Canonical Transformation's Lost Twin Sister'', \textbf{PIRSA:2510.0166}
	\item  Oct 27: The Planimeter and the Weyl-Heisenberg Group of Kinematic Transformations, \textsection\ref{Oct27} | \textbf{PIRSA:2510.0008}
	\item  Oct 30: The Planimeter and the Weyl-Heisenberg Group of Canonical Transformations, \textsection\ref{Oct30} | \textbf{PIRSA:2510.0205}
	\item  Nov 06: Measuring Instruments, Indirect Measurement, and Positive Gaussian Kraus Operators, \textsection\ref{Nov06} | \textbf{PIRSA:2511.0086}
	\item  Nov 10: Gaussian Channels, Gaussian Random Unitaries, and Sequential Measurements, \textsection\ref{Nov10} | \textbf{PIRSA:2511.0011}
	\item  Nov 13: Wiener Increments and the It\^o Rules, \textsection\ref{Nov13} | \textbf{PIRSA:2511.0095}
	\item  Nov 20: General Gaussian Kraus Operators, Lindblad Operators, and Diffusive Measurement, \textsection\ref{Nov20} | \textbf{PIRSA:2511.0096}
	\item  Nov 24: The Markov Property and The Central Limit Theorem, \textsection\ref{Nov24} | \textbf{PIRSA:2511.0015}
	\item  Nov 27: Instrumental Groups, von Neumann Measurement, and Diffusive Heterodyne, \textsection\ref{Nov27} | \textbf{PIRSA:2511.0097}
	\item  Dec 04: The Stratonovich Product, Maurer-Cartan Stochastic Differential, and Indirect Homodyne, \textsection\ref{Dec04} | \textbf{PIRSA:2512.0018}
	\item  Dec 08: The Kraus-Operator Density and The Fokker-Planck-Kolmogorov Equation, \textsection\ref{Dec08} | \textbf{PIRSA:2512.0015}
	\item  Dec 11:  Noncommutative Simultaneous Measurements and The Instrument Manifold Program, \textsection\ref{Dec11} | \textbf{PIRSA:2512.0019}
\end{enumerate}

Lectures were given every Monday and Thursday except on Nov 03, Nov 17, and Dec 01 when there was no class.

%\section{Introduction: Diffusive Measurements and the Instrument Manifold Program}

%``No Instrumentation without Implementation''?

%The meaning of the term ``universal''.

%The concept of an Instrumental Lie Group, transformations versus operations.

%The Instrument Manifold Program began with the discovery that the spin-coherent-state POVM could be implemented practically and universally by the simultaneous diffusive measurement of the three orthogonal components of angular momentum.

%The original goal of this course was to understand how informationally complete quantum tomography works for the two most fundamental types of physical systems, those consisting of canonically conguate pairs (either position \& momentum or bosonic field quadratures) and those made of angular momentum (or spin for short).\\

\subsection*{References by the Author}

\begin{enumerate}
	\item ``Optimal Pure-State Qubit Tomography via Sequential Weak Measurements'', \textit{Shojaee, Jackson, Riofrio, Kalev, and Deutsch}, \textbf{arXiv:1805.01012}
	\item ``How to perform the coherent measurement of a curved phase space by continuous isotropic measurement.  I.  Spin and the Kraus operator geometry of $\mathrm{SL}(2,\mathbb{C})$'', \textit{Jackson and Caves}, \textbf{arXiv:2107.12396}
	\item ``Simultaneous Momentum and Position Measurement and the Instrumental Weyl-Heisenberg Group'', \textit{Jackson and Caves}, \textbf{arXiv:2306.01045}
	\item ``Simultaneous Measurements of Noncommuting Observables: Positive Transformations and Instrumental Lie Groups'', \textit{Jackson and Caves}, \textbf{arXiv:2306.06167}
	\item  ``Sequential Quantum Measurements and the Instrumental Group Algebra'', \textit{Jackson}, \textbf{arXiv:2510.13980}
\end{enumerate}

\pagebreak

\section{Reversible Instruments: The Polar Planimeter and the Weyl-Heisenberg Group}

The general idea was originially from Hermann (1818).

The polar-planimeter in particular was invented by Amsler-Laffon (1854).

Maxwell knew about this as he had also been working on his own model of the planimeter (1854/55).

Courant, student and collaborator of Hilbert's as well as director of Gottingen (before fleeing to New York with John),
refers to the Amsler planimeter in his calculus textbook (1936).

Burkard Polster (a.k.a. The Mathologer) posted a really nice \href{https://www.youtube.com/watch?v=8CksuEk2jJQ&t=740s}{video}  on YouTube on June-07-2025.

\subsubsection{The Canonical One-Form, \textbf{Oct 27}}\label{Oct27}

\noindent (The notes for \textbf{PIRSA:2510.0008} begin here.)\\

\begin{figure}[h!]
	\caption{Picture of the polar planimeter given in class.  Also, watch \textbf{PIRSA:2510.0166} (time stamp 6:11.)}
\end{figure}

Let $\vec{N}_\text{polar}$ be the unit vector field along which a displacement of the tracer would generate rotations of the wheel the fastest.
The change in the angle of the wheel after taking the tracer along a curve $\gamma$ is therefore
\begin{equation}
	\phi = \frac{1}{h}\int_\gamma \vec{N}_\text{polar} \cdot d\vec{r}
\end{equation}
where
\begin{equation}
	h = \text{radius of the meter wheel}.
\end{equation}
Actually, the object being integrated
\begin{equation}
	\lambda_\text{polar} = \vec{N}_\text{polar} \cdot d\vec{r}
\end{equation}
is more naturally and more generally a one-form because it integrates the displacement vectors along the curve (what John Wheeler called ``bongs of the bell''.)
However, for historical reasons it is common among scientists and engineers to formalize contour integrals with ``vector calculus'' notation.
In any case, we will soon enough end up treating the one-form directly.
We will call it the canonical one-form both because it is the very thing that defines how the planimeter works but also because it is analogous to Poincare's canonical one-form in classical mechanics, also called the velocity one-form by Arnold.

\begin{figure}[h!]
	\caption{Diagram of the unit vector field given in class.  Also, see \textbf{PIRSA:2510.0166} (time stamp 6:11.)}
\end{figure}

\subsubsection{Deriving the Canonical One-form of the Polar Planimeter}

Partially in terms of angular co\"ordinates, the unit vector field can be clearly seen from the diagram to be
\begin{align}
	\vec{N}_\text{polar} \cdot d\vec{r}
	&= \sin(\alpha - \beta)dq + \cos(\alpha - \beta)dp\\
	&= (\sin\alpha\cos\beta - \cos\alpha\sin\beta)dq + (\cos\alpha\cos\beta+\sin\alpha\sin\beta)dp.
\end{align}
Defining half the distance between the tracer and the fixed point to be
\begin{equation}
	a^2 \equiv \frac{q^2+p^2}{4},
\end{equation}
some basic trigonometry can be read off the diagram as well:
\begin{equation}
	-\sin\beta = \frac{q}{2a}
	\hspace{50pt}
	\text{,}
	\hspace{50pt}
	\cos\beta = \frac{p}{2a}
	\hspace{50pt}
	\text{,}
\end{equation}
\begin{equation}
	\sin\alpha = \frac{a}{R} 
	\hspace{50pt}
	\text{, and}
	\hspace{50pt}
	\cos\alpha = \frac{1}{R}\sqrt{R^2 - a^2}
\end{equation}
where $R$ is the length of either arm, which we are assuming are equal for simplicity.

The canonical one-form in Cartesian co\"ordinates is therefore
\begin{align}\label{master}
	\lambda_\text{polar}
	&\equiv \vec{N}_\text{polar}\cdot d\vec{r}\\
	&= \left(\frac{a}{R} \frac{p}{2a} +  \frac{1}{R}\sqrt{R^2 - a^2}\frac{q}{2a}\right)dq
	+ \left( \frac{1}{R}\sqrt{R^2 - a^2}\frac{p}{2a} - \frac{a}{R} \frac{q}{2a}\right)dp\\
	&= \left(\frac{p}{2R} +  \frac{q}{2Ra}\sqrt{R^2 - a^2}\right)dq
	+ \left( \frac{p}{2Ra}\sqrt{R^2 - a^2} - \frac{q}{2R}\right)dp\\
	&= \left(\frac12 p +  \frac{q}{2a}\sqrt{R^2 - a^2}\right)dq
	+ \left( \frac{p}{2a}\sqrt{R^2 - a^2} - \frac12 q\right)dp\\
	&= \frac{p\,dq - q\,dp}{2} + \frac{1}{2a}\sqrt{R^2 - a^2}(p \, dp + q \, dq)\\
	&= \frac{p\,dq - q\,dp}{2} + \frac{1}{a}\sqrt{R^2 - a^2}\,d a^2\\
	&= \frac{p\,dq - q\,dp}{2} + 2\sqrt{R^2 - a^2} da\\
	&= \frac{p\,dq - q\,dp}{2} + dG_\text{polar}\\
	&= p dq + dF_\text{polar}
\end{align}
where
\begin{equation}
	G_\text{polar} = R^2\left(2\theta+\sin2\theta\right)
	\hspace{50pt}
	\text{,}
	\hspace{50pt}
	\sin\theta = \frac{a}{R} = \sqrt{\frac{p^2 + q^2}{4R^2}}
\end{equation}
and
\begin{equation}
	F_\text{polar} = -\frac{qp}{2} + G_\text{polar}.
\end{equation}
The exact expression for $F_\text{polar}$ is not too important for what we are here to show, but it is nonetheless nice to just know what it is.

\subsubsection{The Contact One-Form and Kinematic Transformations}

The configuration space of the planimeter is 3-dimensional, two dimensions to locate the tracer and one more dimension for the angle of the wheel.
All displacements $\overline{X}$ in this space that respect the planimeter must satisfy the constriant of being annihilated
\begin{equation}
	\alpha_\text{polar}(\overline{X})=0
\end{equation}
by the contact one-form
\begin{equation}
	\alpha_\text{polar} \equiv \lambda_\text{polar} - h\,d\phi.
\end{equation}
Such displacements will be called \emph{kinematic} displacements.
In particular, if $X$ is a vector in $(q,p)$-space, then
\begin{equation}
	\overline{X} \equiv X + \frac1h \lambda(X) \partial_\phi
\end{equation}
is a kinematic displacement.

This means if we move the tracer along $\partial_q$ or $\partial_p$, then there must also be an accompanying movement in the wheel, generated by $\partial_\phi$.
It is clear that
\begin{equation}
	\lambda_\text{polar}(\partial_q) = p + F_q
	\hspace{50pt}
	\text{and}
	\hspace{50pt}
	\lambda_\text{polar}(\partial_p) = F_p
\end{equation}
where
\begin{equation}
	F_q \equiv \frac{\partial F_\text{polar}}{\partial q}
	\hspace{50pt}
	\text{and}
	\hspace{50pt}
	F_p \equiv \frac{\partial F_\text{polar}}{\partial p}.
\end{equation}
Therefore the lifted displacements that are annihilated by $\alpha_\text{polar}$ are
\begin{equation}
	h\overline{\partial_q} \equiv h \partial_q + (p+F_q) \partial_\phi
	\hspace{50pt}
	\text{and}
	\hspace{50pt}
	h\overline{\partial_p} \equiv h \partial_p + F_p \partial_\phi.
\end{equation}
These diffeomorphisms do not commute.
Rather, their commutator is
\begin{equation}
	[h\overline{\partial_p},h\overline{\partial_q}] = h\partial_\phi.
\end{equation}
Further, the commutators to second order are all zero,
\begin{equation}
	[\partial_\phi,h\overline{\partial_q}] = [\partial_\phi,h\overline{\partial_p}] = 0.
\end{equation}
Therefore, the physical motions of the planimeter are the displacements of a Weyl-Heisenberg group!

\pagebreak
\subsubsection{Canonical Transformations, \textbf{Oct 30}}\label{Oct30}

\noindent (The notes for \textbf{PIRSA:2510.0205} begin here.)\\

There is in fact another, deeper Weyl-Heisenberg group within the planimeter.\\

To motivate the definition of a Canonical Transformation, recall the Euler-Lagrange equation but express it as
\begin{equation}
	p \equiv \left.\frac{\partial L}{\partial \dot{q}}\right|_{q,\dot{q}}
	\hspace{50pt}
	\text{and}
	\hspace{50pt}
	\dot{p} = \left.\frac{\partial L}{\partial q}\right|_{q,\dot{q}}.
\end{equation}
These equations are equivalent to the one-form equation
\begin{align}
	dL &= \dot{p} \, dq + p \, d\dot{q}\\
		&= \lim_{\epsilon\to\infty}\frac{(p+\epsilon \dot{p})d(q+\epsilon \dot{q})- p \,dq}{\epsilon}.
\end{align}
Notice that the final expression is a kind of derivative.
It is a special case of what is known as the Lie derivative of the one-form $p\,dq$ in the direction $d/dt$.

More generally, the Lie derivative $\L_X$ with respect to the infinitesimal transformation generated by $X$ of a differential form $\lambda(q,p)$ on a two-dimensional manifold is
\begin{equation}
	\L_X[\lambda] = \lim_{\epsilon\to0} \frac{\lambda(q+\epsilon X[q], p+\epsilon X[p])-\lambda(q,p)}{\epsilon}.
\end{equation}
The Lie derivative can of course be defined for any number of dimensions.
In particular
\begin{equation}
	\L_X[f dg] = X[f]dg + f d X[g].
\end{equation}

Check that the Cartesian displacements of the planimeter tracer generate canonical transformations:
\begin{equation}
	\L_{\partial_q}[\lambda_\text{polar}] = dF_q
	\hspace{50pt}
	\text{and}
	\hspace{50pt}
	\L_{\partial_p}[\lambda_\text{polar}] = d\left(q+F_p\right)
\end{equation}
where the functions in their Lagrangians are shorthand for the partial derivatives,
\begin{equation}
	F_q = \frac{\partial F_\text{polar}}{\partial q}
	\hspace{50pt}
	\text{and}
	\hspace{50pt}
	F_p = \frac{\partial F_\text{polar}}{\partial p}.
\end{equation}

\subsubsection{Contact Transformations}

If $\alpha$ is a contact one-form, a tangent vector $X$ is said to generate a \emph{contact transformation} if
\begin{equation}
	\L_X[\alpha] = Z \alpha
\end{equation}
for some function $Z$.
A contact transformation is said to be \emph{strict} if $Z=0$.
Canonical transformations of $\lambda_\text{polar}$ are equivalent to strict contact transformations of $\alpha_\text{polar}$ because if $\L_X[\lambda_\text{polar}] = dL$ then $\L_X[\alpha_\text{polar}] = 0$ where
\begin{equation}
	\widehat{X} \equiv X + \frac1h L \partial_\phi.
\end{equation}
Show this.

Therefore the two vector fields
\begin{equation}
	\widehat{\partial_q} = \partial_q + \frac{1}{h}F_q\partial_\phi
	\hspace{50pt}
	\text{and}
	\hspace{50pt}
	\widehat{\partial_p} = \partial_p + \frac1h \left(q+F_p\right)\partial_\phi
\end{equation}
generate (strict) contact transformations | that is
\begin{equation}
	\L_{\widehat{\partial_q}}[\alpha] = \L_{\widehat{\partial_p}}[\alpha] = 0.
\end{equation}

Show that these generators have first-order Lie bracket
\begin{equation}
	[h\widehat{\partial_p},h\widehat{\partial_q}] = -h\partial_\phi
\end{equation}
and second-order Lie brackets
\begin{equation}
	[\partial_\phi,\widehat{\partial_p}] = [\partial_\phi,\widehat{\partial_p}] = 0.
\end{equation}
This establishes that the polar planimeter is a perfect embodiment of the Weyl-Heisenberg group!\\

\subsubsection{Hamiltonians}

Observe that for any contact transformation $\widehat{X}$
\begin{align}
	\alpha(\widehat{X})
	&\equiv \alpha(X + \frac1h L \partial_\phi)\\
	&= \alpha\Big(-\frac1h \lambda(X)\partial_\phi + \frac1h L \partial_\phi\Big)\\
	&= \big(\lambda(X) - L\big)\alpha(-\partial_\phi)\\
	&= \lambda(X) - L.
\end{align}
This can be recognized as a generalization of the Legendre transformation!
For the standard Legendre transform, just consider $\lambda = p\, dq$ and $X = d/dt$.
This means $\alpha(\widehat{X})$ is the Hamiltonian!

In particular the Hamiltonians of the canonical displacements are
\begin{equation}
	\alpha(\widehat{\partial_q}) = p
	\hspace{50pt}
	\text{,}
	\hspace{50pt}
	\alpha(\widehat{\partial_p}) = -q
	\hspace{50pt}
	\text{, and}
	\hspace{50pt}
	\alpha(-\partial_\phi) = h1.
\end{equation}

\pagebreak
\section{Quantum Measuring Instruments}

\subsection{The system-meter interaction model and Kraus operators, \textbf{Nov 06}}\label{Nov06}

\noindent (The notes for \textbf{PIRSA:2511.0086} begin here.)\\

\begin{figure}[h!]
	\caption{Diagram of the system-meter interaction.  Drawn in class.}
\end{figure}

Indirect measurements are modeled by a system-meter interaction, where the system of interest is not measured directly but rather interacts with a meter and then it is the meter that is measured directly.

We will be mostly interested in measurements of meter position, where we imagine a real number, $x$, (rather than an integer) is registered to within some arbitrarily small but not infinite precision $dx \ni x$.

The meter consists of a ``pointer'' that prepared in a normalized initial pure state, $\ket{0}_\text{M}$.
The interaction between the meter and the system is a joint unitary transformation, $\U_\text{int}$.
The meter then terminates in a measurement that registers the real c-number `$x$' represented by the linear functional $\sqrt{dx}\bra{x}_\text{M}$.

\subsubsection{Comments on the registration of a continuous variable}

In analogy to separable bases, an expression like
\begin{equation}
	\text{``}\int_\R dx \proj{x}_M = 1_\text{M} \text{''}
\end{equation}
is often written down.
However, it should be kept in mind that $\ket{x}_\text{M}$ will not need to exist as a physical state for our calculations.
Indeed, such an infinitely precise state is considered impossible to prepare.
As such, the prepared initial pointer state $\ket{0}_\text{M}$ must not be confused with the (impossible to prepare and therefore nonexistent) ``state'' vector $\ket{x=0}_\text{M}$.

On the other hand, the linear functional $\bra{x}_\text{M}$ is required to exist for our calculations.
This brings up a subtle point about what it means to register `$x$' because it is equally impossible to store such a number in memory.
However, this subtle detail is managed by $dx$.

For further details about the mathematical details of directly measuring position, I recommend reading arXiv:2510.13980 [page 14, section II.A] by me, ``The role of the rigged Hilbert space in Quantum Mechanics'' (2005) by Rafael de la Madrid, and ``Probabilistic and Statistical Aspects of Quantum Theory'' by Alexander Holevo.\\

That said and done, the essense of it all is that we will need the operators $Q_\text{M}$ and $\frac{1}{\hbar}P_\text{M}$ defined by the properties
\begin{equation}
	\bra{x}_M Q_\text{M} = x\bra{x}_M
	\hspace{50pt}
	\text{and}
	\hspace{50pt}
	\bra{x}_M \frac{1}{\hbar}P_\text{M} = -i\frac{\partial}{\partial x}\bra{x}_M
\end{equation}
which clearly satsify the Weyl-Heisenberg canonical commutation relation,
\begin{equation}
	[Q_\text{M},P_\text{M}]=i1_\text{M}.
\end{equation}

\subsubsection{Back to indirect measurement}

Returning to the indirect measurement, the probability of registering `$x$' within the infinitesimal region $dx$ is, according to the Born rule,
\begin{align}
	dx P(x) &= dx\sum_{\phi}\sum_{\psi}p_\psi \Big|\bra{\phi}_\text{S}\bra{x}_\text{M}\U_\text{int}\ket{\psi}_\text{S}\ket{0}_\text{M}\Big|^2.
\end{align}
The ``average over initial states and sum over final states" $\sum_\psi p_\psi \sum_\phi$ is a mantra that field theorists are usually taught very early in their training.

A tool that is very useful for conceptualizing this calculation is the so-called \emph{Kraus operator}
\begin{equation}
	\Omega_x \equiv \sqrt{dx} K_x \equiv \sqrt{dx} \bra{x}_\text{M}\U_\text{int}\ket{0}_\text{M},
\end{equation}
where I haven't come up with a good name yet for the ``Kraus operator'' $K_x$ without the square-root of the measure in it.

With this tool, we can massage the probability to register `$x$' into the form
\begin{align}
	dx P(x) &= \sum_{\phi}\sum_{\psi}p_\psi \Big|\bra{\phi}_\text{S}\Omega_x\ket{\psi}_\text{S}\Big|^2\\
	 &= \sum_{\phi}\sum_{\psi}p_\psi \bra{\phi}_\text{S}\Omega_x\ket{\psi}_\text{S}\bra{\psi}_\text{S}\Omega_x^\dag\ket{\phi}_\text{S}\\
	 &= \sum_{\phi} \bra{\phi}_\text{S}\Omega_x\left(\sum_{\psi}p_\psi\ket{\psi}\bra{\psi}_\text{S}\right)\Omega_x^\dag\ket{\phi}_\text{S}\\
	 &= \tr_\text{S} \!\big(\Omega_x \rho \Omega_x^\dag\big)
\end{align}
where
\begin{equation}
	\rho \equiv \sum_{\psi}p_\psi\ket{\psi}\bra{\psi}_\text{S}
\end{equation}
is the usual (von Neumann), density operator, in this case normalized such that $\tr_\text{S} \rho = 1$.

Note that by construction, the Kraus operators satsify a type of \emph{completeness relation},
\begin{equation}\label{complete}
	\sum_x \Omega_x^\dag \Omega_x = \int_\R dx\, K_x^\dag K_x = 1_\text{S}.
\end{equation}

\subsubsection{General definitions}

A \emph{generalized measurement} is any set of operators $\{\Omega_x\}$ which satisfy the completeness relation (equation \ref{complete}).
The elements of a generalized measurement are called \emph{measurement operators}, \emph{back actions}, or \emph{Kraus operators}.
The index $x$ is called the \emph{register}.
Upon registering $x$, a state $\rho$ is understood to update to transition to the state $\Omega_x\rho\Omega_x^\dag$.
There is a bit of flexibility as to where and when the state is normalized.
When simulating state evolution, it is more efficient to always renormalize the state, thus making the update rule nonlinear.
When solving for the generalized mesurements that result from sequences of generalized measurements, one must keep the linear update rule.

Every generalized measurement defines a \emph{generalized observables} or \emph{positive-operator-valued measure}, usually called a \emph{POVM}, which is the set of positive operators $\{\Omega_x^\dag \Omega_x\}$ such that the completeness relation (equation \ref{complete}) is satisfied.
The POVM is less information than the generalized measurement because it does not specify how the states are updated.
In particular, the generalized measurements $\{U_x\Omega_x\}$ all define the same POVM for any set of unitaries $\{U_x\}$.
The elements of a POVM are called \emph{effects} or more often just \emph{POVM elements}.
Given a POVM $\{E_x\}$, the \emph{generalized Luders-von Neumann rule} is to assume the generalized measurement consists of its positive square-roots, $\{\sqrt{E_x}\}$.
Therefore, any generalized measurement with positive Kraus operators will be called a \emph{Luders measurement}.\\

It is also very useful to consider the update rule $\rho \mapsto \Omega_x \rho \Omega_x^\dag$ more explicitly as a superoperator,
\begin{equation}
	\O_x \equiv \Odot[K_x] \equiv K_x \odot K_x^\dag
\end{equation}
where $\odot$ denotes a tensor product with the implied action $A\odot B^\dag (\rho) \equiv A \rho B^\dag$.
Superoperators such as $\Odot[K_x]$ which represent physical state updates are generally called \emph{operations}.
The set of operations $\{O_x\}$ which satisfy the completeness relation (equation \ref{complete}) is called an \emph{instrument}.
The elements of an instrument are called \emph{selective operations} or just \emph{instrument elements}.
The selective operations we consider here, $\Odot[K_x]$, are said to be \emph{Kraus-rank-one}.
Generally, the selective operations are positive mixtures of Kraus-rank-one operations.
Such mixtures represent imperfections in the instrument.

Often the terms ``instrument'', ``generalized measurement'', and just ``measurement'' will be used interchangably.
Similarly, the terms ``selective operation'', ``instrument element'', and ``Kraus operator'' will often be used interchangably.
Of course, one should always understand the difference between an operator and a superoperator, but otherwise the distinct terminology can be relaxed in certain circumstances.

Given an instrument $\{\O_x\}$, the superoperator
\begin{equation}
	\Z \equiv \sum_x \O_x
\end{equation}
is called the \emph{total operation}, the \emph{nonselective operation}, the \emph{decoherence channel}, or just the \emph{channel}, and often haphazardly just \emph{the operation}.
The channel has two defining properties and is said to be \emph{completely positive} or ``\emph{CP}'' and \emph{trace-preserving} or ``\emph{TP}''.

\subsection{The Gaussian Luders Instrument}

The meter preparation we will be most interested in is the Gaussian pure state, with wavefunction
\begin{equation}
	\sqrt{dx}\braket{x}{0}_\text{M} = \sqrt{\frac{dx}{\sqrt{2\pi \sigma^2}}e^{-x^2/2\sigma^2}}.
\end{equation}
The two most important examples of such a meter preparation are the center-of-mass position of an ion sent through a Stern-Gerlach apparatus and a mode of the electric field in vacuum.
Such a state is most efficiently defined by the linear, first-order differential equation
\begin{equation}
	\left(\frac{x}{2\sigma}+\sigma\frac{\partial}{\partial x}\right)\braket{x}{0}_\text{M} = 0,
\end{equation}
which is equivalent to the more abstract annihilation property
\begin{equation}
	a_\text{M}\ket{0}_\text{M} = 0
	\hspace{50pt}
	\text{where}
	\hspace{50pt}
	a_\text{M} \equiv \frac{1}{2\sigma}Q_\text{M} + i\frac{\sigma}{\hbar}P_\text{M}.
\end{equation}

The first system-meter interaction we will consider is the \emph{controlled displacement},
\begin{equation}
	\U_\text{int} = e^{-i\frac{\sigma}{\hbar} P_\text{M}\otimes 2X_\text{S} \theta}
\end{equation}
where the parameter $\theta$ is either proportional to the interaction time or square-root of the interaction time, depending on depending on whether the interactions are ballistic or diffusive.
More about this will be explained when we get to decoherence.
In the meantime, we can calculate the Kraus operators:
\begin{align}
	K_x &\equiv \bra{x}_\text{M}\U_\text{int}\ket{0}_\text{M}\\
	& = e^{-2 X_\text{S} \theta \sigma \partial_x}\braket{x}{0}_\text{M}\\
	& = \sqrt{\frac{1}{\sqrt{2\pi\sigma^2}} e^{-\frac{1}{2\sigma^2}(x-2\sigma\theta X_\text{S})^2}}\\
	& = \frac{1}{(2\pi\sigma^2)^{1/4}} e^{-\frac{1}{4\sigma^2}(x-2\sigma\theta X_\text{S})^2}\\
	& = \frac{1}{(2\pi\sigma^2)^{1/4}} e^{-\theta^2(\frac{x}{2\sigma \theta}- X_\text{S})^2} \label{SG}\\
	& = \sqrt{\frac{1}{\sqrt{2\pi\sigma^2}} e^{-x^2/2\sigma^2}}e^{\frac{x}{\sigma}\theta X_\text{S} - \theta^2 X_\text{S}^2}\label{Diff}.
\end{align}
Since the Kraus-operators are positive Hermitian, this instrument is Luders.

These Kraus operators are the first examples of the ``universal'' expressions I emphasized during the introduction of this class.
What ``universal'' means in this case is that these Kraus operators, their completeness relation, their POVM, and their decoherence channel are already defined without having specified anything about the system observable, $X_\text{S}$, such as its spectrum or even the Hilbert space it acts on.
Because of this universal nature, these Kraus operators are said to represent \emph{positive transformations}, which I will explain more soon.

In the case of the Stern-Gerlach experiment, $X_\text{S} = J_z$.
This is still considered a universal expression because all of the instrument-related quantities still make sense without specifying the Hilbert space, which in this case means specifying the quantum spin number $j$.
By saying ``$X_\text{S} = J_z$'', what we are actually doing is adding the commutation relations for rotation into the mix.

Now suppose the spectrum \emph{is} specified and that the spectrum is discrete (i.e. separable).
Let $\lambda$ be an eigenvalue of $X_\text{S}$ and $\Pi_\lambda$ be the corresponding orthogonal projector so that
\begin{equation}
	X_\text{S} = \sum_\lambda \lambda \, \Pi_\lambda.
\end{equation}
From expression \ref{SG} we see that, regardless of the state of the system, the Kraus operator $K_x$ will go to zero in the limit of a strong interaction $\theta \gg 1$ unless $x = 2\sigma \theta \lambda$ for some eigenvalue $\lambda$.

\pagebreak

\subsection{The Gaussian Luders Channel, \textbf{Nov 10}}\label{Nov10}

\noindent (The notes for \textbf{PIRSA:2511.0011} begin here.)\\

Recall the Kraus operators for the controlled displacement of a pure Gaussian meter,
\begin{align}
	K_x = \sqrt{\frac{1}{\sqrt{2\pi\sigma^2}} e^{-x^2/2\sigma^2}}e^{\frac{x}{\sigma}\theta X - \theta^2 X^2}
\end{align}
where now I'm dropping the `system' subscript, $X = X_\text{S}$.
We can easily calculate their corresponding decoherence channel,
\begin{align}
	\Z &= \int dx \; K_x \odot K_x^\dag\\
	&= e^{- \theta^2 X^2} \odot e^{- \theta^2 X^2} \circ \int \frac{dx}{\sqrt{2\pi\sigma^2}} e^{-x^2/2\sigma^2}\; e^{\frac{x}{\sigma}\theta X}\odot e^{\frac{x}{\sigma}\theta X}\\
	&= e^{- \theta^2 (X^2\odot 1 + 1 \odot X^2)} \circ \int \frac{dx}{\sqrt{2\pi\sigma^2}} e^{-x^2/2\sigma^2}\; e^{\frac{x}{\sigma}\theta (X\odot 1 + 1 \odot X)}\\
	&= e^{- \theta^2 (X^2\odot 1 + 1 \odot X^2)} \circ e^{\frac12 \sigma^2 \left(\frac{1}{\sigma}\theta (X\odot 1 + 1 \odot X)\right)^2}\\
	&= e^{\theta^2 \left(-X^2\odot 1 - 1 \odot X^2 +\frac12 (X\odot 1 + 1 \odot X)^2\right)}\\
	&= e^{-\theta^2 \frac12 (X\odot 1 - 1 \odot X)^2}\\
	&= e^{-\theta^2 \frac12 \ad_X^2}
\end{align}
where
\begin{equation}
	\ad_X \equiv X\odot 1 - 1 \odot X
	\hspace{50pt}
	\text{or equivalently}
	\hspace{50pt}
	\ad_X(\rho) = [X,\rho].
\end{equation}
This particular decoherence channel is also called the Feynman-Vernon influence functional because it is the single-mode analog of the indirect measurements of a quantum field they considered in 1963.

\subsection{Random Gaussian Unitaries and an Important Symmetry of the Channel}

If we change the interaction to a unitary where instead the meter is ``boosted'' by the system observable $Y_\text{S}$,
\begin{equation}
	\U_\text{int} = e^{i\frac{1}{\sigma} Q_\text{M}\otimes Y_\text{S} \theta},
\end{equation}
then the Kraus operator (which is quite a bit easier to calculate) becomes
\begin{align}
	K_x = \sqrt{\frac{1}{\sqrt{2\pi\sigma^2}} e^{-x^2/2\sigma^2}}e^{i\frac{x}{\sigma} \theta Y}.
\end{align}
This instrument generates exactly the same decoherence channel as the Luders instrument,
\begin{align}
	\Z &= \int \frac{dx}{\sqrt{2\pi\sigma^2}} e^{-x^2/2\sigma^2}\; e^{i\frac{x}{\sigma}\theta Y}\odot e^{-i\frac{x}{\sigma}\theta Y}\\
	&= \int \frac{dx}{\sqrt{2\pi\sigma^2}} e^{-x^2/2\sigma^2}\; e^{i\frac{x}{\sigma}\theta (Y\odot 1 - 1 \odot Y)}\\
	&= e^{-\theta^2 \frac12 (Y\odot 1 - 1 \odot Y)^2}\\
	&= e^{-\theta^2 \frac12 \ad_Y^2}.
\end{align}
On the other hand, it generates the completely uninformative POVM,
\begin{align}
	\Omega_x^\dag \Omega_x = \frac{dx}{\sqrt{2\pi\sigma^2}} e^{-x^2/2\sigma^2}1_\text{S}.
\end{align}

\subsection{Sequential Measurements}

If we repeat the measurement $\mathcal{I} \equiv \{\Omega_x\}$ $n$ times, then we can consider this sequence as a composite instrument with Kraus operators
\begin{equation}
	\Omega_{(x_1,x_2,\ldots,x_n)} = \Omega_{x_n}\cdots \Omega_{x_1}.
\end{equation}
This new instrument can be expressed as $\mathcal{I}^{*n} \equiv \mathcal{I}*\mathcal{I}*\cdots *\mathcal{I}$, where $\mathcal{B}*\mathcal{A}$ denotes the free product,
\begin{equation}
	\mathcal{B}*\mathcal{A} \equiv \Big\{BA : B \in \mathcal{B} \;\;\&\;\; A \in \mathcal{A}\Big\}.
\end{equation}
So for a sequence of two instruments, we could also write for the composite instrument
\begin{equation}
	\{\Omega_y\}_y*\{\Upsilon_x\}_x = \{\Omega_y\Upsilon_x\}_{(x,y)}.
\end{equation}
I will eventually call this free-product a convolution, but probably not until next season.

Given two instruments, $\mathcal{A} = \{\Upsilon_x\}_x$ and $\mathcal{B} = \{\Omega_y\}_y$, the channel of their sequential instrument is therefore the superoperator composition of their individual channels,
\begin{align}
	\Z_{\mathcal{B}*\mathcal{A}}
	&= \sum_{x,y} \Odot[\Omega_y\Upsilon_x]\\
	&= \sum_{x,y} \Odot[\Omega_y]\circ\Odot[\Upsilon_x]\\
	&= \sum_y \Odot[\Omega_y]\circ\sum_x \Odot[\Upsilon_x]\\
	&= \Z_{\mathcal{B}}\circ\Z_{\mathcal{A}}
\end{align}
where remember $\Odot[K]\equiv K \odot K^\dag$ and $A \odot B^\dag (\rho) \equiv A \rho B^\dag$.
In particular, for a repeated instrument
\begin{align}
	\Z_{\mathcal{I}^{*n}} = (\Z_{\mathcal{I}})^{\circ n}.
\end{align}\\

If $\mathcal{I}$ is one of the Gaussian instruments of the previous section, then
\begin{equation}
	\Z_{\mathcal{I}^n} = e^{-n\theta^2 \frac12 \ad_X^2}.
\end{equation}
This means that if $t$ is the time it takes to perform $\mathcal{I}$ a single time, then
\begin{equation}
	\theta = \sqrt{\kappa t}
\end{equation}
for some constant $\kappa$, with units of inverse time, called the \emph{measurement rate}.

\pagebreak
\section{Diffusive Quantum Measurements}

\subsection{Some Stochastic Calculus for Diffusive Markov Processes, \textbf{Nov 13}}\label{Nov13}

\noindent (The notes for \textbf{PIRSA:2511.0095} begin here.)\\

\subsubsection{The Wiener Increment}

If the Gaussian instrument is weak, then
\begin{equation}
	\theta = \sqrt{\kappa dt}.
\end{equation}
In this case, the random variable for the register is instead expressed in terms of the standard Wiener increment,
\begin{align}
	dW \equiv \frac{x}{\sigma}\sqrt{dt},
\end{align}
in which case we write, for example, the Kraus operators of the Gaussian Luders instrument as
\begin{align}
	\Omega_x
	&\equiv \sqrt{\frac{dx}{\sqrt{2\pi\sigma^2}}e^{-x^2/2\sigma^2}}
	e^{\frac{x}{\sigma}\sqrt{\kappa dt} X - \kappa dt X^2}\\
	&= \sqrt{\frac{d(dW)}{\sqrt{2\pi dt}}e^{-dW^2/2dt}}
	e^{\sqrt{\kappa} dW X - \kappa dt X^2}\\
	&\equiv \sqrt{d\mu(dW)} K_{dW}
\end{align}
where I define the Wiener increment measure and the conjugate Kraus operator,
\begin{align}
	d\mu(dW) \equiv \frac{d(dW)}{\sqrt{2\pi dt}}e^{-dW^2/2dt}
	\hspace{50pt}
	\text{and}
	\hspace{50pt}
	K_{dW} \equiv e^{\sqrt{\kappa} dW X - \kappa dt X^2}.
\end{align}

\subsubsection{Diffusive records and It\^o rules}

A diffusive measurement record over a time interval $[0,T)$ is a sequence of $T/dt$ independent Wiener increments which we denote by
\begin{equation}
	dW_{[0,T)} \equiv \{dW_0,dW_{dt},dW_{2dt},\ldots,dW_{T-dt}\}.
\end{equation}
The measure of this sequence is called the Wiener-path-integral measure
\begin{align}
	\D\mu\left[dW_{[0,T)}\right]
	&\equiv \prod_{n=0}^{T/dt-1} d\mu(dW_{n dt})\\
	&= e^{-\sum_{n=0}^{T/dt-1} (dW_{ndt})^2/2dt} \prod_{n=0}^{T/dt-1} \frac{d(dW_{n dt})}{\sqrt{2\pi dt}}\\
	&\equiv e^{-\int_{t=0}^{T_-} (dW_t)^2/2dt} \prod_{n=0}^{T/dt-1} \frac{d(dW_{n dt})}{\sqrt{2\pi dt}}
\end{align}
where $T_- \equiv T - dt$.

I will denote the average of any functional of the Wiener path against the Wiener path-integral measure by angle brackets,
\begin{align}
	\Big\langle f[dW_{[0,T)}]\Big\rangle \equiv \int \D\mu[dW_{[0,T)}]\,f[dW_{[0,T)}]
\end{align}

When doing calculus with these increments, various forms of integration occur which in the limit $dt \to 0$ amount to what is called the It\^o rule.
The It\^o rule is a bit subtle, so rather than state it generally, I will introduce it first through a set of special cases.\\

Essentially, it is already clear by definition that
\begin{align}
	\langle dW_s dW_t\rangle = \delta_{st} dt
\end{align}
where $\delta_{st}$ is a Kronecker delta and not a Dirac delta.
(In certain instances, it can be useful to think of the Dirac delta as $\delta(s-t) = \delta_{st}/dt$.)
However, in the limit $dt \to 0$, what is also true is that the fluctuations of the random variable $dW_s dW_t$ never actually add up.
This means that in the limit $dt \to 0$, we can use the much stronger It\^o rule,
\begin{align}
	dW_s dW_t = \delta_{st} dt
\end{align}
as I will now explain.

\paragraph{$dW_t^2 = dt$}:  Specifically, the fluctuations of the random variable $dW_t^2$ can be neglected in the limit $dt \to 0$.

To see this, consider the random variable
\begin{equation}
	x \equiv \int_{t=0}^{T_-} dW_t^2.
\end{equation}
For its first moment, we have
\begin{align}
	\langle x \rangle
	&= \int_{t=0}^{T_-} \langle dW_t^2 \rangle \\
	&= \int_{t=0}^{T_-} dt\\
	&= T.
\end{align}
For its second moment, we have
\begin{align}
	\langle x^2 \rangle
	&= \int_{s=0}^{T_-} \int_{t=0}^{T_-} \langle dW_s^2 dW_t^2 \rangle\\
	&= \int_{t=0}^{T_-} \langle dW_t^4 \rangle + 2\int_{s=0}^{T_-} \int_{t=0}^{s_-} \langle dW_s^2\rangle \langle dW_t^2 \rangle\\
	&= \int_{t=0}^{T_-} 3dt^2 + 2\int_{s=0}^{T_-} \int_{t=0}^{s_-} ds dt\\
	&= 3Tdt + T^2.
\end{align}
Therefore, the variance is
\begin{align}
	\langle x^2 \rangle - \langle x \rangle^2 = 3 T dt,
\end{align}
which is indeed negligible in the limit $dt \to 0$.\\

\paragraph{$dW_t dt = 0$}:

Consider the random variable
\begin{align}
	x \equiv \int_{t=0}^{T_-} dW_t dt.
\end{align}
Clearly
\begin{align}
	\langle x \rangle &= \int_{t=0}^{T_-} \langle dW_t \rangle dt\\
	&=0.
\end{align}
Further
\begin{align}
	\langle x^2 \rangle &= \int_{s=0}^{T_-} \int_{t=0}^{T_-} \langle dW_s dW_t\rangle (dt)^2\\
	&= \int_{s=0}^{T_-} \int_{t=0}^{T_-} \delta_{st} (dt)^3\\
	&= \int_{t=0}^{T_-} (dt)^3\\
	&= T (dt)^2.
\end{align}\\

\paragraph{$dW_t dV_t = 0$}:  Specifically, the time-wise product of any two independent Wiener paths is effectively zero.

Consider the random variable
\begin{equation}
	x \equiv \int_{t=0}^{T_-}  dW_t dV_t.
\end{equation}

Clearly
\begin{align}
	\langle x \rangle = 0.
\end{align}
Further
\begin{align}
	\langle x^2 \rangle
	&= \int_{s=0}^{T_-}  \int_{t=0}^{T_-}  \langle dW_s dV_s dW_t dV_t \rangle\\
	&= \int_{s=0}^{T_-}  \int_{t=0}^{T_-}  \langle dW_s dW_t\rangle \langle dV_s dV_t \rangle\\
	&= \int_{s=0}^{T_-}  \int_{t=0}^{T_-}  \delta_{st} dt \delta_{st} dt\\
	&= \int_{s=0}^{T_-}  \int_{t=0}^{T_-}  \delta_{st} (dt)^2\\
	&= Tdt.
\end{align}

\pagebreak

\subsection{Diffusive Measurements, \textbf{Nov 20}}\label{Nov20}

\noindent (The notes for \textbf{PIRSA:2511.0096} begin here.)\\

\subsubsection{Weak Gaussian Measurements in General and Diffusive Instruments}

For weak Gaussian measurements, we can calculate the Kraus operators for the more general interaction that includes both types considered,
\begin{align}
	\U_\text{int}
	&= e^{-i(\frac{2\sigma}{\hbar} P_\text{M}\otimes X_\text{S}-\frac{1}{\sigma} Q_\text{M}\otimes Y_\text{S})\sqrt{\kappa dt}}\\
	&= e^{(a^\dag_\text{M}\otimes L_\text{S} - a_\text{M}\otimes L^\dag_\text{S})\sqrt{\kappa dt}}
\end{align}
where recall the annihilator of the meter state is
\begin{align}
	a_\text{M} \equiv \frac{1}{2\sigma}Q_\text{M} + i\frac{\sigma}{\hbar}P_\text{M}
\end{align}
but also introduced is the system operator
\begin{equation}
	L_\text{S} \equiv X_\text{S} + i Y_\text{S}.
\end{equation}

Dropping the system subscripts, a careful calculation reveals two very useful expressions for the Kraus operators,
\begin{align}
	\Omega_{dW} &= \sqrt{d\mu(dW)}e^{L \sqrt{\kappa}dW-\frac12(L^\dag L + L^2)\kappa dt}\\
	&= e^{iY\sqrt{\kappa}dW - i\frac12 \{X,Y\}\kappa dt}\sqrt{d\mu(dW)}e^{X \sqrt{\kappa}dW-X^2\kappa dt}
\end{align}
where $\{X,Y\} \equiv XY + YX$ is the anticommutator.
Examing the second expression, one can see that it is the Hermitian component of $L$ that is being measured while anti-Hermitian component of $L$ generates a locally adaptive unitary as a well as introducing a nonadaptive unitary drift term generated by the anticommutator which we will see in terms of phase spaces (planar and spherical) always has the effect of squeezing the quasiprobability of the measured state.

For a good exercise, calculate these expressions for the Kraus oeprator.
It requires a few versions of the Ito rule, namely $dW^2 = dt$ and $dW dt = 0$, which I will explain in a moment.
For assistance, you can look to arXiv:2510.13980, Appendix A.2, page 39.\\

Further, a calculation reveals the decoherence channel to be
\begin{align}
	\Z &= e^{(L\odot L^\dag - \frac12 L^\dag L \odot 1 - \frac12 1 \odot L^\dag L)\kappa dt}.
\end{align}
The superoperator
\begin{equation}
	\D_L \equiv L\odot L^\dag - \frac12 L^\dag L \odot 1 - \frac12 1 \odot L^\dag L
\end{equation}
is known as a \emph{dissipator} or \emph{Lindbladian}, which means that this operator introduced, $L$, is in fact a \emph{Lindblad operator}.

The Lindbladian generator is the most general form of a continuous-in-time decoherence channel.
However, the Gaussian instrument which registers a so-called \emph{diffusive process} (a.k.a. \emph{Wiener process} or \emph{Brownian motion}) isn't the only instrument that generates this type of decoherence.
Namely, there are also the instruments that register \emph{jump processes} such as indirect photodetection which can also be thought of as measurements of a qubit meter.
Diffusive and jump processes are the primary examples of continuous Markov processes.
In this course, the focus is almost entirely on instruments with diffusive registers.

\subsubsection{Diffusively Measuring NonCommuting Observables Simultaneously}

Consider two Wiener measurements with Lindblad operators $L$ and $M$, with Kraus operators
\begin{align}
	\Upsilon_{dV} = \sqrt{d\mu(dV)}e^{M \sqrt{\kappa}dV-\frac12(M^\dag M + M^2)\kappa dt}
	\hspace{25pt}
	\text{and}
	\hspace{25pt}
	\Omega_{dW} = \sqrt{d\mu(dW)}e^{L \sqrt{\kappa}dW-\frac12(L^\dag L + L^2)\kappa dt}.
\end{align}
An essential thing to understand is that even if the Lindblad operators do not commute, the Kraus operators they generate actually do still commute because, of the It\^o rules.
In particular, 
\begin{align}
	\Upsilon_{dV}\Omega_{dW}
	&= \Omega_{dW} \Upsilon_{dV} e^{[M,L]\kappa dV dW}\\
	&= \Omega_{dW} \Upsilon_{dV}.
\end{align}

Crucially, what this means is that noncommuting observables can be diffusively measured simultaneously.
Given a tuple of Lindblad operators, $\vec{L} = (L_1,L_2,\ldots,L_n)$, it therefore makes sense to consider the diffusive measurement generated by the  Kraus operators
\begin{align}
	\Omega_{d\vec{W}} = \sqrt{\frac{d^n(d\vec{W})}{(2\pi dt)^{n/2}}e^{-d\vec{W}^2/2dt}}e^{\vec{L} \cdot \sqrt{\kappa}d\vec{W} - Q_{\vec{L}}\kappa dt}
\end{align}
where $d\vec{W} = (dW^1, dW^2, \ldots, dW^n)$ is the conjugate tuple of independent Wiener records, and the infinitesimal generator consists of the stochastic linear component
\begin{align}
	\vec{L} \cdot d\vec{W} \equiv L_1 dW^1 + \ldots + L_n dW^n
\end{align}
and a deterministic quadratic (drift) component,
\begin{align}
	Q_{\vec{L}} = \frac12 \sum_{i=1}^n L_i^\dag L_i + L_i^2.
\end{align}

An abbreviation I sometimes use for these measurements is DMNCOS (pronounced ``demon cause'') which stands for \emph{diffusively measuring noncommuting observables simultaneously}.
However, I tend to just say \emph{simultaneous measurements} in my day-to-day.
If a simultaneous measurement is continuously repeated for a time $T$, then the instrument will register a record of Wiener increments $d\vec{W}_{[0,T)}$ corresponding to the Kraus operator
\begin{align}
	\Omega_{d\vec{W}_{[0,T)}} = \sqrt{\D\mu\!\left[d\vec{W}_{[0,T)}\right]}\,\mathcal{T}\exp \left\{\int_0^{T_-}\vec{L} \cdot \sqrt{\kappa}d\vec{W}_t - Q_{\vec{L}}\kappa dt\right\}
\end{align}
where the (stochastic) time-ordered exponential has been introduced,
\begin{align}
	\mathcal{T}\exp \left\{\int_{t=0}^{T_-} \updelta_t \right\}
	\equiv e^{\updelta_{T-dt}}e^{\updelta_{T-2dt}} \cdots e^{\updelta_{dt}}e^{\updelta_0}.
\end{align}

\subsubsection{The fundamental problem of continuous measurement}

The fundamental problem of continuous measurement is whether an instrument such as $\{\Omega_{d\vec{W}_{[0,T)}}\}$ can be automated, i.e. calculated with a classical computer.
In this case, the instrument can be considered an automaton.
One general instance where the instrument is an automaton is when the time-ordered exponentials universally generate a finite-dimensional group.
This is determined by considering the span of all the commutators between the Lindblad generators and the quadratic term to all orders.
If the time-ordered exponentials generate a finite-dimensional group, then the instrument is said to be a \emph{principal instrument}.
Otherwise, the exponentials generate an infinite-dimensional group and we say we have a \emph{chaotic instrument}.
All principal instruments can be considered automata thanks to the theory of finite-dimensional Lie groups.
Whether there are \emph{chaotic instruments} that can also be automated or not is actually an open question.
I suspect there are.
For anyone curious, I recommend they try reading arXiv:2306.06167 or arXiv:2510.13980.

In general, I call this group generated by the time-ordered exponential the \emph{instrumental group}.
There are 4 particularly important principal instruments that I've studied quite a bit, only 2 of which I have so far written about (extensively).
We will discuss them a bit next semester as they are the path to the phase space POVMs, planar, spherical, and beyond.
For now, let me just introduce them a little here:\\

\paragraph{The Isotropic Spin Measurement (ISM), $\vec{L} = (J_x,J_y,J_z)$}:

This measurement generates a 7-dimensional instrumental group I call $\mathrm{ISpin}(3,\C) \cong \R\times\SL(2,\C)$.
For a quick read from when we first discovered this, read arXiv: 1805.01012.
For an extensive mathematical exploration of how ISM realizes the spin-coherent POVMs universally, read arXiv:2107.12396.\\

\paragraph{SPQM, $\vec{L} = (Q,P)$}:

The acronym stands for \emph{Simultaneous momentum (p) and position (q) measurement}, but it is also an homage to Antoine Barchielli at the University of Milan who first attempted to solve this problem in 1981.
SPQM is also a rif on the Milan extension of the emblematic SPQR or \emph{Senatus Populesque Romanus}.

This measurement also generates a 7-dimensional instrumental group I call the instrumental Weyl-Heisenberg group, $\mathrm{IWH}$.
For an extensive mathematical exploration of how SPQM realizes the standard coherent POVM, read arXiv:2306.01045\\

\paragraph{Indirect heterodyne, $\vec{L} = (a, -ia)$}:

This measurement generates a 3-dimensional instrumental group which is far easier to navigate.
It was first discovered and analyzed in 1994 in two independent papers, by Goetsch \& Graham and by Wiseman \& Milburn.

Like SPQM, indirect heterodyne realized the standard coherent POVM.
The essential difference between these measurements is that SPQM leaves the post-measurement state in a random coherent state while indirect heterodyne leaves the post-measurement state in vacuum.\\

\paragraph{Indirect spin-heterodyne, $\vec{L} = (J_+, -iJ_+, J_z)$}:

This measurement generates a 3-dimensional instrumental group which is also much easier to navigate.
Wiseman and Milburn got close to finding this measurement in 1995.
As Howard Wiseman put it when I explained it to him in 2022, ``Who knew a third measurement could make the problem infinitely simpler?!''.

Like ISM, indirect spin-heterodyne realizes the standard coherent POVM.
The essential difference between these measurements is that ISM leaves the post-measurement state in a uniformly random spin-coherent state while indirect heterodyne leaves the post-measurement state in the spin-coherent state pointing in the positive z-direction.\\

\begin{figure}
	\caption{Insert a diagram of the isotropic spin-measurement (drawn in class.)}
\end{figure}

\begin{figure}
	\caption{Insert a diagram of indirect heterodyne (drawn in class.)}
\end{figure}

\pagebreak

\subsection{Why we Need the It\^o Rules: The Central Limit Theorem at Work, \textbf{Nov 24}}\label{Nov24}

\noindent (The notes for \textbf{PIRSA:2511.0015} begin here.)\\

Another property that the term ``universal'' is refering to is the \emph{dynamic universality class} of Markov processes, traditionally known as the \emph{Central Limit Theorem}.

\subsubsection{Discovering the need to keep $\epsilon^2$ from Weak Binary POVMs}

What I was calling $\theta$ at the beginning of this course I am now calling $\epsilon$ for Rob Spekkens.\\

Consider the two-outcome weak POVM with elements
\begin{align}
	E_b \equiv \frac12 \left(1 + 2b\epsilon X\right)
\end{align}
where $b\in\{+1,-1\}$ and $\epsilon \ll 1$.
One might think this means the Kraus operators of the Luders instrument are
\begin{align}
	\Omega_b \overset{?}{=} \sqrt{\frac12}\left(1+b\epsilon X\right).
\end{align}
However, this would imply that the channel is
\begin{align}
	\Z &\equiv \sum_b \Omega_b \odot \Omega_b^\dag\\
	&= \sum_b \frac12 \left(1\odot 1 + b\epsilon(X\odot 1 + 1 \odot X) + \epsilon^2 b^2 X \odot X\right)\\
	&= 1\odot 1 +  \epsilon^2 X \odot X.
\end{align}
This tells us two things.  The first is that once again, the decoherence time must be proportional to $\epsilon^2$.
The second is that, since we must therefore keep around terms up to $\epsilon^2$, there is an additional term that must be in the weak Luders Kraus-operator, else the channel would not be CPTP:
\begin{align}
	\Omega_b = \sqrt{\frac12}\left(1+b\epsilon X - \frac12 \epsilon^2 X^2\right).
\end{align}
Amazingly, the POVM element still has no term of order $\epsilon^2$,
\begin{align}
	\Omega_b^\dag \Omega_b
	\equiv \frac12\left(1+2b\epsilon X +\O(\epsilon^3)\right).
\end{align}

\subsubsection{Discovering the Gaussian Instrument from the Weak Sequential Binary Instrument}

Having sorted out the Kraus operators of a weak two-outcome measurement, a sequence of such measurements would have Kraus operators
\begin{align}
	\Omega_{b_n} \cdots \Omega_{b_1}
	&=\sqrt{\frac{1}{2^n}}\left(1+\epsilon (b_1+\ldots+b_n)X + \epsilon^2 \sum_{i\neq j}b_i b_j X^2- \frac12 n \epsilon^2 X^2\right).
\end{align}
This double sum is too cumbersome and can be avoided if we instead express our Kraus operators in terms of exponentials,
\begin{align}
	\Omega_b = \sqrt{\frac12} e^{\epsilon b X - \epsilon^2 X^2},
\end{align}
in which case
\begin{align}
	\Omega_{b_n} \cdots \Omega_{b_1} = \sqrt{\frac{1}{2^n}}e^{\epsilon (b_1+\ldots+b_n)X - n \epsilon^2 X^2}.
\end{align}

There are clearly several distinct sequences of $(b_1,\ldots,b_n)$ that give the same total Kraus operator, $\binom{n}{k}$ of them to be exact, where $k$ is the number of $b_i = +1$.
More convenient than the counting parameter $k$ is just the sum of the binomial trials,
\begin{align}
	m
	&\equiv b_1 + \ldots + b_n\\
	&= 2k - n.
\end{align}
With this, we collect all the distinct sequences which give the same total Kraus operator.
Recall $\Odot[\Omega] \equiv \Omega \odot \Omega^\dag$ and define $\Omega_m$ such that
\begin{align}
	\Odot\left[\Omega_m\right]
	&\equiv \sum_{b_1+\ldots + b_n = m} \Odot\left[\Omega_{b_n} \cdots \Omega_{b_1}\right]\\
	&=\frac{1}{2^n} \sum_{b_1+\ldots + b_n = m} \Odot\left[e^{\epsilon (b_1+\ldots+b_n)X - n \epsilon^2 X^2}\right]\\
	&=\frac{1}{2^n} \binom{n}{\frac{n+m}{2}} \Odot\left[e^{\epsilon m X - n \epsilon^2 X^2}\right]
\end{align}
or back in terms of just the Kraus operator,
\begin{align}
	\Omega_m =  \sqrt{\frac{1}{2^n}\binom{n}{\frac{n+m}{2}}}e^{\epsilon m X - n \epsilon^2 X^2}
\end{align}
As $n$ increases, the binomial distribution very quickly approximates a Gaussian distribution
\begin{align}
	\frac{1}{2^n}\binom{n}{k} &\sim  \frac{1}{\sqrt{2\pi (n/4)}}e^{-\frac{1}{2(n/4)}(k-\frac{n}{2})^2}\\
	&= \frac{2}{\sqrt{2\pi n}} e^{-m^2/2n}.
\end{align}
For a finite, but large enough $n$, say $n=10$, because these weak sequential random variables become effectively Gaussian, we can substitute them for standard Wiener increments by defining
\begin{align}
	\kappa dt \equiv n \epsilon^2 
	\hspace{25pt}
	\text{,}
	\hspace{25pt}
	\sqrt{\kappa} dW \equiv m \epsilon
	\hspace{25pt}
	\text{, and therefore}
	\hspace{25pt}
	\sqrt{\kappa}d(dW) = 2 \epsilon.
\end{align}
Under this co\"ordinate transformation, the Kraus operators are therefore
\begin{align}
	\Omega_m &\sim \sqrt{\frac{2}{\sqrt{2\pi n}} e^{-\frac{m^2}{2n}}}e^{\epsilon m X - n \epsilon^2 X^2}\\
	&= \sqrt{\frac{2 \epsilon }{\sqrt{2\pi n \epsilon^2}} e^{-\frac{(m \epsilon)^2}{2n\epsilon^2}}}e^{\epsilon m X - n \epsilon^2 X^2}\\
	&= \sqrt{\frac{\sqrt{\kappa}d(dW)}{\sqrt{2\pi \kappa dt}} e^{-\frac{(\sqrt\kappa dW)^2}{2\kappa dt}}}e^{\sqrt\kappa dW X - \kappa dt X^2}\\
	&= \sqrt{\frac{d(dW)}{\sqrt{2\pi dt}} e^{-dW^2/2 dt}}e^{\sqrt\kappa dW X - \kappa dt X^2},
\end{align}
which are exactly the weak Gaussian Kraus operators we have been discussing.\\

This is one of the simplest (and historically the oldest [de Moivre, 1733]) examples of the central limit theorem which in turn is the simplest example of a dynamic universality class.

\subsubsection{Back to It\^o}

So if we consider weak Gaussian Kraus operators with Hermitian Lindblad operators,
\begin{align}
	\Omega^{(X)}_{dW} \equiv \sqrt{\frac{d(dW)}{\sqrt{2\pi dt}} e^{-dW^2/2 dt}}e^{\sqrt\kappa dW X - \kappa dt X^2},
\end{align}
the subtle point is that, by Baker-Campbell-Hausdorff
\begin{align}
	\Omega^{(X)}_{dW}\Omega^{(Y)}_{dV}
	&= \sqrt{\frac{d(dW) d(dV)}{2\pi dt} e^{-(dW^2+dV^2)/2 dt}}
	e^{\sqrt\kappa (dW + dV) X + \frac12 [X,Y] dW dV - \kappa 2dt X^2 + \O(dt^{3/2})}\\
	&= \Omega^{(Y)}_{dV}\Omega^{(X)}_{dW}e^{[X,Y] dW dV + \O(dt^{3/2})}
\end{align}
and therefore one needs to understand, for example, why $dWdV=0$ | that is, why certain fluctuating terms of order $dt$ don't matter (i.e. make no contribution to the finite-time stochastic variables), even though non-fluctuating terms of order $dt$ clearly do matter.

\pagebreak

\section{The Instrumental Group: Working with the Time-Ordered Exponential Universally (Nov 27)}\label{Nov27}

\noindent (The notes for \textbf{PIRSA:2511.0097} begin here.)\\

Recall that the Kraus operators of a diffusive measurement are always of the form
\begin{align}
	\Omega_{d\vec{W}_{[0,T)}} = \sqrt{\D\mu\!\left[d\vec{W}_{[0,T)}\right]}\,\mathcal{T}\exp \left\{\int_0^{T_-}\vec{L} \cdot \sqrt{\kappa}d\vec{W}_t - Q_{\vec{L}}\kappa dt\right\}
\end{align}
where the quadratic generator is
\begin{align}
	Q_{\vec{L}} = \frac12 \sum_{i=1}^n L_i^\dag L_i + L_i^2.
\end{align}
and the Lindblad operators, of course, do not have to commute.

This section is a set of examples to help familiarize ourselves with the time-ordered exponential.

\subsection{The Universal Property of the Time-Ordered Exponential}

The brownian motion defined by the set of Lindblad operators,
\begin{align}
	\gamma\left[d\vec{W}_{[0,T)}\right]
	\equiv \mathcal{T} e^{\int_{t=0}^{T_-}\updelta(d\vec{W}_t)},
\end{align}
is \emph{universal} in the sense that it is defined by just the abstract commutators of the generators in $\updelta(d\vec{W}_t)$ and holds true for \emph{all} representations thereof.
The span of all the commutators of the generators in $\updelta(d\vec{W}_t)$ is a Lie algebra and the time-ordered exponentials and their brownian motion are said to reside in the so-called \emph{universal covering group} of this Lie algebra.

In the context of diffusive measurements, I call this universal covering group the \emph{Instrumental Group} and denote it by $\mathrm{IG}$.

One could define an instrumental semigroup,
\begin{equation}
	\bigcup_{T \in [0,\infty)}\left\{\mathcal{T} e^{\int_{t=0}^{T_-}\updelta(d\vec{W}_t)}\right\}_{d\vec{W}_{[0,T)}} \subset \mathrm{IG}
\end{equation}
where remember that the time-ordered exponential is
\begin{align}
\mathcal{T} e^{\int_{t=0}^{T_-}\updelta_{t}}
&\equiv \mathcal{T} \prod_{k=0}^{T_-/dt}e^{\updelta_{k dt}}\\
&= e^{\updelta_{T_-}} e^{\updelta_{T_- - dt}} \cdots e^{\updelta_{2dt}} e^{\updelta_{dt}} e^{\updelta_0}.
\end{align}

\subsection{Example 1: Diffusive Measurement of a Single Hermitian Observable and von Neumann Measurement}\label{indecisive}

The Kraus operators for indirectly measuring a single observable diffusively (by an infinite number of weak von Neumann interactions) are
\begin{align}
	\Omega^{\text{Diff}}_{dW_{[0,T)}}
	&= \sqrt{\D\mu\left[dW_{[0,T)}\right]}\mathcal{T}
	e^{\int_0^{T_-} X \sqrt{\kappa} dW_t - X^2 \kappa dt}\\
	&= \sqrt{\D\mu\left[dW_{[0,T)}\right]}
	e^{X \sqrt{\kappa} \int_0^{T_-} dW_t - X^2 \kappa T}.
\end{align}
For a single Hermitian observable, we can collect all the Kraus-operators that are equal, to find
\begin{align}
	\Odot\left[\Omega^{(X)}_{W_T} \right]
	&\equiv \int_{\int_0^{T_-} dW_t = W_T} \Odot\left[\Omega^{(X)}_{dW_{[0,T)}}\right]\\
	&= \int_{\int_0^{T_-} dW_t = W_T} \D\mu\left[dW_{[0,T)}\right]\,\Odot\left[e^{X \sqrt{\kappa} W_T - X^2 \kappa T}\right]\\
	&= d(W_T)\int \D\mu\left[dW_{[0,T)}\right] \delta\!\left(W_T - \int_0^{T_-} dW_t\right)\,\Odot\left[e^{X \sqrt{\kappa} W_T - X^2 \kappa T}\right]\\
	&= \frac{d(W_T)}{\sqrt{2\pi T}}e^{-W_T^2/2T}\,\Odot\left[e^{X \sqrt{\kappa} W_T - X^2 \kappa T}\right].
\end{align}
Therefore
\begin{align}
\Omega^{\text{Diff}}_{W_T}
&= \sqrt{\frac{d(W_T)}{\sqrt{2\pi T}}e^{-W_T^2/2T}}\,e^{X \sqrt{\kappa} W_T - X^2 \kappa T}\\
&= \sqrt{\frac{d(W_T)}{\sqrt{2\pi T}}e^{-\frac{1}{2T}(W_T - 2\sqrt{\kappa} T X)^2}}\\
&= \sqrt{\frac{du}{\sqrt{2\pi}}e^{-\frac{1}{2}(u - 2\sqrt{\kappa T} X)^2}}.
\end{align}\\

The most interesting part of the above is showing
\begin{align}
\int \D\mu\left[dW_{[0,T)}\right] \delta\!\left(W_T - \int_0^{T_-} dW_t\right)
= \frac{1}{\sqrt{2\pi T}}e^{-W_T^2/2T}.
\end{align}
We will do this by showing that the path-integral satisfies a Fokker-Planck-Kolmogorov forward equation which in this case is just the standard heat equation.
This is the simplest example of a Feynman-Kac formula.

Let
\begin{align}
	D_T(x) \equiv \int \D\mu\left[dW_{[0,T)}\right] \delta\!\left(x - \int_0^{T_-} dW_t\right).
\end{align}
Then
\begin{align}
	D_{T+dt}(x) 
	&\equiv \int \D\mu\left[dW_{[0,T+dt)}\right] \delta\!\left(x - \int_0^{(T+dt)_-} dW_t\right)\\
	&= \int d\mu(dW_T)\int \D\mu\left[dW_{[0,T)}\right] \delta\!\left(x - dW_T -  \int_0^{T_-} dW_t\right)\\
	&= \int d\mu(dW_T) D_T(x-dW_T)\\
	&= \int d\mu(dW_T) e^{-dW_T \frac{\partial}{\partial x}}[D_T](x)\\
	&=  e^{\frac12 dt \frac{\partial^2}{\partial x^2}}[D_T](x)\\
	&=  D_T(x) + \frac12 dt \frac{\partial^2}{\partial x^2}[D_T](x)
\end{align}
and therefore
\begin{align}
	\frac{\partial}{\partial t}D_t(x) = \frac12 \frac{\partial^2}{\partial x^2}[D_t](x).
\end{align}\\

These diffusive Kraus operators are quite different from the Kraus operators for indirectly measuring a single observable by a single von Neumann interaction,
\begin{align}
	\Omega^\text{vN}_{x}
	&\equiv \sqrt{dx} \bra{x}_{\text{M}} e^{-i \frac{2\sigma}{\hbar} P_\text{M} \otimes X gT} \ket{0}_{\text{M}}\\
	&= \sqrt{\frac{dx}{\sqrt{2\pi \sigma^2}} e^{-(x-2\sigma gT X)^2/2\sigma^2}}\\
	&= \sqrt{\frac{du}{\sqrt{2\pi}} e^{-(u-2 gT X)^2/2}}.
\end{align}

\subsubsection*{Eigenstate Collapse (a.k.a. von Neumann Measurement)}

See arXiv: , page ???.

\subsection{Example 2: Diffusive Heterodyne, Our First Noncommutative Principal Instrument}\label{Ex2}

In indirect heterodyne, the Lindblad operators are
\begin{equation}
	\vec{L} = \left(\frac{1}{\sqrt2}a,-\frac{i}{\sqrt2}a\right)
\end{equation}
where the factors of $\sqrt2$ are just a convention.
With these Lindblad operators, the factors of $a^2$ in the quadratic term of the Weak Gaussian Kraus operators therefore cancel and one is left with
\begin{equation}
	\Omega_{d\omega} = \sqrt{d\mu(d\omega)}K_{d\omega},
\end{equation}
where defined are the complex Weiner measure of the increment
\begin{equation}
	d\mu(d\omega) = \frac{d^2(d\omega)}{\pi dt}e^{d\omega^*d\omega/dt},
\end{equation}
and the conjugate Kraus operators
\begin{equation}
	K_{d\omega} = e^{-\frac12 a^{\!\dag} \!a\, \kappa dt+a \sqrt{\kappa} d\omega^*}.
\end{equation}
In this complex notation, the diffusive Kraus operators are 
\begin{align}
	\Omega_{d\omega_{[0,T)}} &= \mathrm{T}\prod_{t=0}^{T/dt-1}\Omega_{d\omega_t}\\
	& = \sqrt{\mathcal{D}\mu \left[d\omega_{[0,T)}\right]}\;\mathrm{T} \exp\left(\int_0^T -\frac12 a^{\!\dag} \!a\, \kappa dt+a \sqrt{\kappa} d\omega_t^*\right),
\end{align}
where the complex Wiener path measure,
\begin{equation}
	\mathcal{D}\mu\!\left[d\omega_{[0,T)}\right] = \prod_{t=0}^{T/dt-1}d\mu(d\omega_t).
\end{equation}\\

The ordered products of these Kraus operators are particular easy to solve because the only noncommutativity left boils down to the identity
\begin{equation}\label{renormalizationW}
	\boxed{
		\vphantom{\Bigg(}
		\hspace{15pt}
		e^{a\, d\omega^*}e^{-\frac12 a^{\!\dag} \!a\, \kappa T}
		= e^{-\frac12 a^{\!\dag} \!a\, \kappa T}e^{a\,d\omega^*e^{-\frac12\kappa T}}.
		\hspace{10pt}
		\vphantom{\Bigg)}
	}
\end{equation}
Applying equation~\ref{renormalizationW}, the time-ordered exponential is easily solved,
\begin{align}
	\Omega_{d\omega_{[0,T)}}
	& = \sqrt{\mathcal{D}\mu \left[d\omega_{[0,T)}\right]}\;
	e^{-\frac12 a^{\!\dag} \!a\, \kappa T}e^{a\,\mu[d\omega_{[0,T)}]^* },
\end{align}
where defined is the fundamental linear functional of the heterodyne process
\begin{equation}\label{heterofunc}
	\mu\!\left[d\omega_{[0,T)}\right]=\int_0^T \!\!\sqrt{\kappa} d\omega_te^{-\frac12\kappa t}.
\end{equation}
Notice how the POVM ceases after a time equal to just a few $1/\kappa$.

\pagebreak
\section{The Stratonovich Product and the Modified Maurer-Cartan Stochastic Differential (Dec 04)}\label{Dec04}

\noindent (The notes for \textbf{PIRSA:2512.0018} begin here.)\\

To solve these time-ordered exponentials more generally, we are going to need a more systematic approach.
This means introducing a couple more tools.

\subsection{The Stratonovich Product}

Because terms of both order $dW$ and order $dW^2=dt$ contribute to stochastic integrals, products of stochastic random variables with stochastic differentials become a bit more subtle.
In accordance with the It\^o rule, the usual rules of differential calculus, the chain rule and the product rule, become modified.
If $x_t$ is a stochastic random variable, then $f_t \equiv f(x_t)$ is a stochastic random variable for any twice-differentiable function $f$, with stochastic differential
\begin{align}
	df_t = f'(x_t) dx_t + \frac12 f''(x_t)dx_t^2.
\end{align}
If $f_t$ and $g_t$ are stochastic random variables, then their product is a stochastic random variable with stochastic differential
\begin{align}
	d(f_t g_t) = (df_t) g_t + f_t dg_t + (df_t)dg_t.
\end{align}

Both of these rules can be made to look just like the chain rule and product rule of differential calculus if we define the appropriate modification to the product between a stochastic random variable and a stochastic differential.
The \emph{Stratonovich product} of a stochastic differential $dg_t$ on the left of a stochastic random variable $f_t$ is defined to be
\begin{align}
	dg_t \circledS f_t \equiv dg_t(f_t + \frac12 df_t),
\end{align}
and similarly for the differential on the right,
\begin{align}
	f_t \circledS dg_t \equiv (f_t + \frac12 df_t)dg_t.
\end{align}
In terms of these, the modified rules of stochastic differential calculus then take a form similar to ordinary differential calculus,
\begin{align}
	df_t = f'(x_t) \circledS dx_t
\end{align}
and
\begin{align}
	d(f_t g_t) = df_t \circledS g_t + f_t \circledS dg_t.
\end{align}

It is also important to know (and easy to show) that the Stratonovich product is associative in the sense that
\begin{align}
	f_t \circledS (g_t \circledS dh_t) = (f_t g_t) \circledS dh_t.
\end{align}
I will leave showing this as an exercise to the reader.

\subsection{The Modified Maurer-Cartan Stochastic Differential}

Consider $dW_{[0,T)}$ fixed and let $x_t$ be the Brownian path in the instrumental group defined by
\begin{align}
	x_t \equiv \gamma\left[dW_{[0,t)}\right].
\end{align}
First, understand that the definition of the time-ordered exponential is equivalent to the sequence of equations
\begin{align}
	x_{t+dt} = e^{\updelta_t}x_t.
\end{align}
As a differential equation, this means
\begin{align}
	dx_t &\equiv x_{t+dt} - x_t\\
	& = (e^{\updelta_t} - 1)x_t\\
	& = (\updelta_t + \frac12 \updelta_t^2)x_t\\
	& = \updelta_t(x_t + \frac12 \updelta_t x_t)\\
	& = \updelta_t \circledS x_t
\end{align}
and therefore
\begin{align}\label{MMCSD}
	dx_t \circledS x_t^\inv = \updelta_t.
\end{align}
The object on the left of this equation I've called the \emph{modified Maurer-Cartan stochastic differential} (MMCSD) since [arXiv: 2107.12396].
However, connecting to the ``Stratonovich picture'' is something I only got around to in [arXiv: 2306.06167].

\subsection{Example 3: Indirect Homodyne, using the new tools}\label{Ex3.1}

In indirect homodyne we once again have just a single Lindblad operator
\begin{align}
	L = \frac{1}{\sqrt2} a
\end{align}
where $a$ is a standard bosonic annihilation operator.
Therefore the infinitesimal generator of the diffusive measurement is
\begin{align}\label{homogen}
	\updelta_t = -L^\dag L \frac12 \kappa dt - L^2 \frac12 \kappa dt + L \sqrt\kappa dW_t.
\end{align}
The 3 factors of $\updelta_t$ close under Lie brackets, in particular
\begin{align}
	[L^\dag L, L] = - \frac12 L
	\hspace{50pt}
	\text{and}
	\hspace{50pt}
	[L^\dag L, L^2] = - L^2,
\end{align}
and therefore the time-ordered exponential
\begin{equation}
	\gamma\!\left[dW_{[0,T)}\right] \equiv \mathcal{T}e^{\int_0^{T_-}\updelta_t}
\end{equation}
resides in a 3-dimensional instrumental group.

In particular, the elements of this group can always be put in the form
\begin{equation}
	x = e^{-L^\dag L r} e^{-L^2 s} e^{L q}
\end{equation}
as I will show inductively by translating the time-ordered exponential into stochastic differential equations for the coordinates $(r,s,q)$.\\

\subsubsection*{Deriving the Coordinate SDEs}

If we accept the decomposition
\begin{align}
	x = e^{-L^\dag L r} e^{-L^2 s} e^{L q},
\end{align}
we can use it to transform infinitesimal coordinate displacements to right-invariant displacements as I will now show.
Observe that if we differentiate with respect to $r$, holding $s$ and $q$ constant,
\begin{align}
	\partial_r[x]
	&= \partial_r[e^{-L^\dag L r}] e^{-L^2 s} e^{L q}\\
	&= -L^\dag L e^{-L^\dag L r} e^{-L^2 s} e^{L q}\\
	&= -L^\dag \! L\, x.\label{RinvEx1}
\end{align}
A little more interesting is if we differentiate with respect to $s$, holding $r$ and $q$ constant,
\begin{align}
	\partial_s[x]
	&= e^{-L^\dag L r} \partial_r[e^{-L^2 s}] e^{L q}\\
	&= e^{-L^\dag L r} (-L^2 e^{-L^2 s}) e^{L q}\\
	&= -e^{-L^\dag L r} L^2 e^{L^\dag L r} x\\
	&= -e^r L^2 x.\label{RinvEx2}
\end{align}
Similarly, differentiating with respect to $s$, holding $r$ and $q$ constant, gives
\begin{align}
	\partial_q[x] = e^{r/2} L x.\label{RinvEx3}
\end{align}

Expanding the stochastic differential in the coordinate basis by the chain rule and dropping the time subscript
\begin{align}
	dx
	&= dr \circledS \partial_r[x] + ds \circledS \partial_s[x] + dq \circledS \partial_q[x]\\
	&= \left(-L^\dag\!L\,dr - L^2 e^r ds + L e^{r/2} dq\right)\circledS x
\end{align}
or equivalently
\begin{align}
	dx \circledS x^\inv
	= -L^\dag\!L\,dr - L^2 e^r ds + L e^{r/2} dq.
\end{align}
Combining this with equations \ref{MMCSD} and \ref{homogen} gives the system of stochastic differential equations
\begin{align}
	\begin{cases}
		dr = \frac12 \kappa dt\\
		e^r ds = \frac12 \kappa dt\\
		e^{r/2} dq = \sqrt\kappa dW
	\end{cases}
\end{align}
which with the initial condition $r(0) = s(0) = q(0) = 0$ is pretty straight-forward to solve.
The trickiest of these is probably the second equation in the system, which with the solution of the first equation is
\begin{align}
	s(T) &= \int_0^{T_-} e^{-\frac12 \kappa t} \frac12 \kappa dt\\
	&= -\int_0^{T_-} de^{-\frac12 \kappa t}\\
	&= 1- e^{-\frac12 \kappa T}.
\end{align}
Altogether, the solution is
\begin{align}\label{filter}
	\begin{cases}
		r_t = \frac12 \kappa t\\
		s_t = 1- e^{-\frac12 \kappa t}\\
		q_t = \sqrt\kappa \int_0^{T_-} e^{-\frac14 \kappa t}dW_t.
	\end{cases}
\end{align}
In particular, we see this same turning off feature in the POVM.
This feature somehow makes the universal instrument far easier to interpret than \ref{indecisive}, where the instrument can't really make up its mind without a spectrum for the observable.

\pagebreak
\section{The Kraus-Operator Density and its Fokker-Planck-Kolmogorov Equation (Dec 08)}\label{Dec08}

\noindent (The notes for \textbf{PIRSA:2512.0015} begin here.)\\

Having some experience with the time-ordered exponential, we now return to the measure component of the diffusive Kraus-operators.
We will be able to collect equivalent Kraus-operators, leaving a Kraus-operator distribution and Kraus-operator density (KOD), similar to what was done in section \ref{indecisive}.
To solve for the KOD generally will require yet another tool, the invariant derivative.

\subsection{Invariant Derivatives, One-Forms, and Integral (Haar) Measures}

Coordinate derivatives are not the natural form of differentiation when working with time-ordered exponentials.
Rather, because the measuring process generates trajectories in the form of equation \ref{MMCSD}, what is natural are the so-called right-invariant derivatives of a group.
Given an element $X$ of the Lie algebra, the right-invariant derivative at a point $x$ in the Lie group of a function $f(x)$ is
\begin{equation}
	\Rinv{X}[f](x) \equiv \lim_{h\to0} \frac{f(e^{hX}x)-f(x)}{h}.
\end{equation}
Similarly, we can take left-invariant derivatives,
\begin{equation}
	\Linv{X}[f](x) \equiv \lim_{h\to0} \frac{f(xe^{hX})-f(x)}{h}.
\end{equation}
``Invariant'' here is really short for ``translation-invariant'', where every element $g$ of the group defines left- and right-translation operators,
\begin{equation}
	\mathfrak{L}_g[f](x) \equiv f(gx)
	\hspace{50pt}
	\text{and}
	\hspace{50pt}
	\mathfrak{R}_g[f](x) \equiv f(xg).
\end{equation}
Therefore the names of the right- and left-invariant derivatives refer to the respective commutativity property
\begin{equation}
	\mathfrak{R}_g \circ \Rinv{X} \circ \mathfrak{R}_g^\inv = \Rinv{X}
	\hspace{50pt}
	\text{and}
	\hspace{50pt}
	\mathfrak{L}_g \circ \Linv{X} \circ \mathfrak{L}_g^\inv = \Linv{X}.
\end{equation}
Finally, on the space of analytic functions the invariant derivatives actually generate the translation operators
\begin{equation}
	\mathfrak{L}_{e^X} = e^{h\Rinv{X}}
	\hspace{50pt}
	\text{and}
	\hspace{50pt}
	\mathfrak{R}_{e^X} = e^{h\Linv{X}}
\end{equation}
which can be considered the noncommutative generalization of Taylor's theorem.\\

Every basis of the Lie algebra defines both a right-invariant basis and a left-invariant basis of vector fields on the universal covering group.
If $\{X_\mu\}$ is a basis of the Lie algebra, the basis of right-invariant one-forms dual to $\{\Rinv{X_\mu}\}$ will be denoted
\begin{equation}
	\theta_{\mathrm{R}}^\mu (\Rinv{X_\nu}) = \delta^\mu_\nu
\end{equation}
and the basis of left-invariant one-forms dual to $\{\Linv{X_\mu}\}$ will be denoted
\begin{equation}
	\theta_{\mathrm{L}}^\mu (\Linv{X_\nu}) = \delta^\mu_\nu.
\end{equation}
The right- and left-invariant integral measures are therefore
\begin{equation}
	d_\mathrm{R}x = |\theta_{\mathrm{R}}^1 \wedge \cdots \wedge \theta_{\mathrm{R}}^n|
	\hspace{50pt}
	\text{and}
	\hspace{50pt}
	d_\mathrm{L}x = |\theta_{\mathrm{L}}^1 \wedge \cdots \wedge \theta_{\mathrm{L}}^n|
\end{equation}
where $n$ is the dimension of the Lie group.

\subsubsection*{Example 3 Again}\label{Ex3.2}

With the right-invariant derivatives at hand, we can rewrite equations \ref{RinvEx1}, \ref{RinvEx2}, and \ref{RinvEx3} as
\begin{align}\label{RinvCoord}
	\begin{cases}
		\partial_r = -\Rinv{L^\dag L}\\
		\partial_s = -e^r \Rinv{L^2}\\
		\partial_q = e^{r/2} \Rinv{L}.
	\end{cases}
\end{align}
To be clear, equations \ref{RinvEx1}, \ref{RinvEx2}, and \ref{RinvEx3} technically say that these derivatives are equal for any position of the IG, but this of course (by the chain rule) implies the derivatives are therefore equal for all functions of position as well.
This means the right-invariant basis in terms of the coordinate basis is
\begin{align}\label{RinvCoordFPKE}
	\begin{cases}
		\Rinv{L^\dag \!L} = -\partial_r\\
		\Rinv{L^2} = -e^{-r}\partial_s\\
		\Rinv{L} = e^{-r/2}\partial_q
	\end{cases}
\end{align}
and we can therefore draw the righ-invariant basis.

\begin{figure}[h!]
	\caption{Picturing the right-invariant vector fields of ${L^\dag}\! L \equiv \frac12 {a^\dag}\! a$ and $L \equiv \frac{1}{\sqrt2} a$.}
\end{figure}

From equations \ref{RinvCoord} we can also immediately read off the right-invariant one-forms,
\begin{equation}
	\begin{cases}
	\theta_\mathrm{R}^{L^\dag \!L} = -dr\\
	\theta_\mathrm{R}^{L^2} = -e^r ds\\
	\theta_\mathrm{R}^{L} = e^{r/2} dq
\end{cases}
\end{equation}
and the right-invariant measure
\begin{align}
	d_\mathrm{R} x
	&= |\theta_\mathrm{R}^{L^\dag \!L}\wedge\theta_\mathrm{R}^{L^2}\wedge\theta_\mathrm{R}^{L}|\\
	&= e^{\frac32 r} dr \, ds \, dq.
\end{align}

\paragraph*{Exercise} Calculate the left-invariant derivatives, dual one-forms, and Haar measure.

Partial answer:
\begin{align}
	d_\mathrm{L} x
	&= dr \, ds \, dq.
\end{align}

\subsection{The Kraus-Operator Distribution, Density, and Fokker-Planck-Kolmogorov Equation}

Recall that the Kraus operators of a diffusive instrument are always of the form
\begin{equation}
	\Omega_{d\vec{W}_{[0,T)}} = \sqrt{\D\mu\left[d\vec{W}_{[0,T)}\right]} K_{\gamma\left[d\vec{W}_{[0,T)}\right]}
\end{equation}
where $\D\mu\left[d\vec{W}_{[0,T)}\right]$ is the Wiener path-integral measure, $\gamma\left[d\vec{W}_{[0,T)}\right]$ is the universal time-ordered exponential, and $K_{yx} = K_y K_x$ is a (tempered) representation of the instrumental group.

Similar to what was done in section \ref{indecisive}, we can collect equivalent Kraus-operators
\begin{equation}
	\Omega_x = \sqrt{d\mu_T(x)} K_{x},
\end{equation}
defining a \emph{Kraus-operator distribution}
\begin{align}
	d\mu_T(x) = \int_{x = \gamma\left[d\vec{W}_{[0,T)}\right]} \D\mu\left[d\vec{W}_{[0,T)}\right].
\end{align}
The Kraus-operator distribution can be easily seen to satisfy a Chapman-Kolmogorov equation
\begin{align}
	d\mu_{t+dt}(x)
	&= \int_{x = \gamma\left[d\vec{W}_{[0,t+dt)}\right]} \D\mu\left[d\vec{W}_{[0,t+dt)}\right]\\
	&= \int d\mu(d\vec{W}_t) \int_{x = e^{\updelta(d\vec{W}_t)}\gamma\left[d\vec{W}_{[0,t)}\right]} \D\mu\left[d\vec{W}_{[0,t)}\right]\\
	&= \int d\mu(d\vec{W}_t) \int_{e^{-\updelta(d\vec{W}_t)}x = \gamma\left[d\vec{W}_{[0,t)}\right]} \D\mu\left[d\vec{W}_{[0,t)}\right]\\
	&= \int d\mu(d\vec{W}_t) \, d\mu_t\left(e^{-\updelta(d\vec{W}_t)}x\right).\label{CK}
\end{align}

Now, if we define the \emph{Kraus-operator density} (KOD) $D_T(x)$ to be the distribution function of the \emph{Kraus-operator distribution} with respect to the \textbf{left-invariant} Haar measure,
\begin{equation}
	d\mu_T(x) \equiv d_\mathrm{L}x \, D_T(x),
\end{equation}
then in particular
\begin{align}
	d\mu_t\left(e^{-\updelta(d\vec{W}_t)}x\right)
	&\equiv d_\mathrm{L}\left(e^{-\updelta(d\vec{W}_t)}x\right) \, D_t\left(e^{-\updelta(d\vec{W}_t)}x\right)\\
	&= d_\mathrm{L}x \, D_t\left(e^{-\updelta(d\vec{W}_t)}x\right).
\end{align}
Returning to equation \ref{CK}, we can drop the left-invariant Haar measure on both sides, leaving
\begin{align}
	D_{t+dt}(x)
	&= \int d\mu(d\vec{W}_t) \, D_t\left(e^{-\updelta(d\vec{W}_t)}x\right)\\
	&= \int d\mu(d\vec{W}) \, D_t\left(e^{-\updelta(d\vec{W})}x\right)\\
	&= \int d\mu(d\vec{W}) \, e^{-\Rinv{\updelta(d\vec{W})}}[D_t]\left(x\right)\\
	&= \int d\mu(d\vec{W}) \, e^{-\sum_i\Rinv{L_i}\sqrt{\kappa}dW^i + \Rinv{Q}\kappa dt}[D_t]\left(x\right)\\
	&= e^{\Rinv{Q}\kappa dt} \prod_i \int d\mu(dW^i) \, e^{-\Rinv{L_i}\sqrt{\kappa}dW^i} [D_t]\left(x\right)\\
	&= e^{\Rinv{Q}\kappa dt} \prod_i e^{\frac12 \Rinv{L_i}\Rinv{L_i}\kappa dt} [D_t]\left(x\right)\\
	&= \left(1+ \left(\Rinv{Q} + \frac12 \sum_i \Rinv{L_i}\Rinv{L_i}\right)\kappa dt\right) [D_t]\left(x\right).
\end{align}
Therefore the KOD satisfies the Fokker-Planck-Kolmogorov (forward) equation (FPKE)
\begin{align}
	\frac1\kappa \frac{\partial}{\partial t} D_t(x) = \left(\Rinv{Q} + \frac12 \sum_i \Rinv{L_i}\Rinv{L_i}\right)[D_t](x).
\end{align}\\

Remember that with the universal Kraus-operator density and an instrumental-group representation $K_x$, the probability of registering the instrumental-group element $x$ after a time $T$ is
\begin{equation}
	\Prob_T\left(x \in d_L x | \rho\right) = d_L x \, D_T(x) \tr \left(K_x^\dag K_x \rho\right).
\end{equation}\\

\subsubsection*{Example 3 One More Time}\label{Ex3.3}

For diffusive homodyne, with equations \ref{RinvCoordFPKE} at hand, the KOD therefore satisfies the FPKE
\begin{equation}
	\frac1\kappa \frac{\partial}{\partial t} D_t = \left( -\frac12 \partial_r - \frac12 e^{-r}\partial_s + \frac12 e^{-r}\partial_q^2\right)[D_t](x).
\end{equation}
It is a straightforward exercise to verify that the solution to this is
\begin{equation}
	D_t\left(e^{-L^\dag L r}e^{-L^2 s} e^{L q}\right) = \delta\left(r - \frac12 \kappa t\right)\delta\left(s -1 + e^{\frac12 \kappa t}\right)\frac{1}{\sqrt{2\pi \sigma^2_t}} e^{-q^2/2\sigma_t^2}.
\end{equation}
where
\begin{equation}
	\sigma_t^2 = 2\kappa (1-e^{-\frac12 \kappa t}).
\end{equation}

The standard deviation of the KOD position variable could also have been calculated from the filter function we solved for in equation \ref{filter},
\begin{align}
	\langle q_T^2\rangle
	&=\kappa \int_0^{T_-} \int_0^{T_-} e^{-\frac14 \kappa s}e^{-\frac14 \kappa t} \langle dW_sdW_t \rangle\\
	&=\kappa \int_0^{T_-} \int_0^{T_-} e^{-\frac14 \kappa s}e^{-\frac14 \kappa t} \delta_{st}dt\\
	&=\kappa \int_0^{T_-}e^{-\frac12 \kappa t}dt\\
	&= 2\kappa (1-e^{-\frac12 \kappa T}).
\end{align}

\pagebreak

\section{Simultaneous Diffusive Measurements of Non-Commuting Observables and the Instrument Manifold Program (Dec 11)}\label{Dec11}

\noindent (The notes for \textbf{PIRSA:2512.0019} begin here.)\\

This lecture explored the three large papers by Jackson and Caves that build the foundation of the Instrument Manifold Program.
\begin{enumerate}
	\item ``How to perform the coherent measurement of a curved phase space by continuous isotorpic measurement.  I.  Spin and the Kraus operator geometry of $\mathrm{SL}(2,\mathbb{C})$''  (\textbf{arXiv:2107.12396})
	\item ``Simultaneous Momentum and Position Measurement and the Instrumental Weyl-Heisenberg Group''  (\textbf{arXiv:2306.01045})
	\item ``Simultaneous Measurements of Noncommuting Observables: Positive Transformations and Instrumental Lie Groups''  (\textbf{arXiv:2306.06167})
\end{enumerate}

Below, are statements of the basic problems of boson quadratures handled in \textbf{arXiv:2306.01045} and spin components handled in \textbf{arXiv:2107.12396}.

\subsection{Example 4: Barchielli and the Simultaneous P \& Q Measurement, ``SPQM''}\label{Ex4}

An oddly more famous simultaneous measurement of $P$ and $Q$ refers to a paper by Arthurs and Kelly from 1965.
The diffusive analog of the original Arthurs-Kelly measurement (where $(X,Y) = (Q,P)$) was first considered in 1982 by Barchielli, et al. and solved in 2023 by Jackson and Caves.
This diffusive analog has been called SPQM as a pun, on the one hand refering to the diffusive ``Simultaneous P \& Q Measurement" and on the other referring to the "Senatus PopulesQue Milan", a derivative of the famous abbreviation SPQR for the Senate and People of Rome, paying homage to Barchielli who happens to be a proud Italian based in Milan.

The Kraus operators of SPQM are
\begin{align}
	\Omega^{(Q,P)}_{d\vec{W}_{[0,T)}} = \sqrt{\D\mu\left[d\vec{W}_{[0,T)}\right]}\mathcal{T}
	e^{\int_0^{T_-} Q \sqrt{\kappa} dW_t^q + P \sqrt{\kappa} dW_t^p - (Q^2 + P^2) \kappa dt}.
\end{align}

This measurement has a 7-real-dimensional instrumental group as can be seen by calculating the commutators between the various components of the infinitesimal generator,
\begin{equation}
	\g = \textrm{span}\{H_o,Q,P,iQ,iP,i1,1\}
\end{equation}
where I've defined the operator
\begin{equation}
	H_o \equiv \frac{Q^2 + P^2}{2}.
\end{equation}\\

For a full analysis of this problem, please see \textbf{arXiv:2306.01045}.

\subsubsection{The Arthurs-Kelly Instrument for Comparison}

Inserting a momentum resolution of the identity,
\begin{align}
	1 = \int_\R dp \proj{p}
	\hspace{50pt}
	\text{where}
	\hspace{50pt}
	\braket{x}{p} = \frac{1}{\sqrt{\pi}}e^{ipx}
\end{align}
the Kraus operators of the Arthurs-Kelly simultaneous measurement are
\begin{align}
	\Omega^\text{AK}_{x,y}
	&\equiv \sqrt{dx \, dy \,} \bra{x}_{\text{M}_1} \bra{y}_{\text{M}_2} e^{-i (P_1 \otimes X + P_2 \otimes Y)gT} \ket{\psi_1}_{\text{M}_1} \ket{\psi_2}_{\text{M}_2}\\
	&= \sqrt{dx \, dy \,} \int_{\R \times \R} \frac{dp \, dq}{2\pi} \psi_1(p)\psi_2(q) e^{ip(x-XgT) + iq(y-YgT)}.
\end{align}

The original Arthurs-Kelly result was about $(X,Y) = (Q,P)$ and the interesting fact that the Kraus operators are the coherent-state projectors at $gT=1$.

\subsection{Example 5: The Isotropic Spin Measurement (ISM)}\label{Ex5}

The universal instrument generated by the Lindblad operators
\begin{align}
	\vec{L} = \left(J_z,J_x,J_y\right)
\end{align}
is principal because the quadratic term
\begin{align}
	Q_{\vec{L}} = J_z^2 + J_x^2 + J_y^2 = \vec{J}^{\:2}
\end{align}
commutes with all the other generators.
The instrumental Lie group is therefore 7-dimensional, with Lie algebra generated by $\{Q_{\vec{L}},J_z,J_x,J_y,iJ_z,iJ_x,iJ_y\}$.

The simple fact that $[J_x,J_y] = iJ_z$ means in the context of diffusive measurements that a product of two non-commuting positive transformations is not generally a positive transformation but rather a positive transformation followed by a unitary.
This simple fact is the basic reason why the problem is so mathematically rich.

This particular measuring process is special because it provided the first practical and universal method for realizing the so-called spin-coherent(-state) POVM.  It was this discovery that started the entire instrument manifold program.\\

For a full analysis of this problem, please see \textbf{arXiv:2107.12396}.

\subsubsection{``Why not Measure Just Two Spin-Components?'': A Chaotic Measuring Instrument}

The universal instrument generated by the Lindblad operators
\begin{align}
	\vec{L} = \left(J_z,J_x\right)
\end{align}
is chaotic because the quadratic term
\begin{align}
	Q_{\vec{L}} = J_z^2 + J_x^2 = \vec{J}^{\:2} - J_y^2 
\end{align}
does not commute with the other generators but rather generates to higher and higher orders arbitrarily large multipoles such as $J_xJ_yJ_z$ and $J_z^3$, etc.

If we only consider the instrument acting on a spin-half system, then the instrument chaos becomes invisible as
\begin{align}
	J_x^2 = J_z^2 = \frac14 1_{2\times 2}.
\end{align}
Of course, this identity is not universal as it is only true for spin-half representations.\\

\pagebreak

\appendix
\section{Exercise 1: The ``Spherimeter'' and the Spin Group}

\subsection{Canonical one-form}

For the spherimeter, the canonical one-form will be
\begin{equation}
	\lambda = \cos\theta \, d\phi
\end{equation}
where $\phi$ and $\theta$ are the usual azimuthal and polar angles of the sphere.
The contact one-form will therefore be
\begin{equation}
	\alpha = \lambda + d\psi
\end{equation}
where $\psi$ is the angle of the turntable.
So the Reeb vector is $\partial_\psi$.

This spherical one-form is not directly analogous to the planar one-form we've considered because the spherical one-form is being considered in polar coordinates while the planar one-form was considered in Cartesian coordinates.
It's good to see how the planar one-form looks in polar coordinates and how it is related to the planar one-form in Cartesian coordinates.
So consider the coordinate transformation
\begin{equation}
	a^2 = \frac{p^2+q^2}{2}
	\hspace{50pt}
	\text{and}
	\hspace{50pt}
	\phi = \arctan \frac{p}{q}
\end{equation}
and observe
\begin{align}
	a^2 d\phi
	& = \frac{p^2+q^2}{2} \frac{p \, dq - q \, dp}{p^2+q^2}\\
	& = \frac{p \, dq - q \, dp}{2}\\
	& = p \, dq  - d\frac{pq}{2}.
\end{align}

\subsection{Kinematic displacements}

Let
\begin{equation}
	\overline{X} = \cos\psi \frac{\partial_\phi - \cos\theta \partial_\psi}{\sin \theta} + \sin\psi \,\partial_\theta,
\end{equation}
\begin{equation}
	\overline{Y} = \cos\psi  \partial_\theta - \sin\psi  \frac{\partial_\phi - \cos\theta \partial_\psi}{\sin \theta},
\end{equation}
and
\begin{equation}
	\overline{Z} = \partial_\psi.
\end{equation}\\

Show that
\begin{equation}
	\alpha(\partial_\phi - \cos\theta \partial_\psi) = \alpha(\partial_\theta) = 0
\end{equation}
and therefore
\begin{equation}
	\alpha(\overline{X}) = \alpha(\overline{Y}) = 0
	\hspace{50pt}
	\text{and}
	\hspace{50pt}
	\alpha(\overline{Z})=1.
\end{equation}\\

Show that
\begin{equation}
	[\overline{Z},\overline{X}] = \overline{Y}
	\hspace{50pt}
	\text{and}
	\hspace{50pt}
	[\overline{Z},\overline{Y}] = -\overline{X}
\end{equation}
and
\begin{equation}
	[\overline{X},\overline{Y}] = \overline{Z}.
\end{equation}\\

\subsection{Canonical displacements}

Consider the Killing fields
\begin{equation}
	X = -\cos\phi \cot \theta \,\partial_\phi - \sin\phi \,\partial_\theta,
\end{equation}
\begin{equation}
	Y  = \cos\phi \,\partial_\theta - \sin\phi \cot \theta \,\partial_\phi,
\end{equation}
and
\begin{equation}
	Z = \partial_\phi.
\end{equation}\\

Show
\begin{equation}
	\L_X \lambda = -d \frac{\cos \phi}{\sin\theta}
	\hspace{50pt}
	\text{and}
	\hspace{50pt}
	\L_Y \lambda = -d \frac{\sin \phi}{\sin\theta}
\end{equation}
and
\begin{equation}
	\L_Z \lambda = 0.
\end{equation}\\

Letting the contact displacements be
\begin{equation}
	\widehat{X} \equiv X + \frac{\cos \phi}{\sin\theta} \partial_\psi
	\hspace{50pt}
	\text{and}
	\hspace{50pt}
	\widehat{Y} \equiv Y + \frac{\sin \phi}{\sin\theta} \partial_\psi
\end{equation}
and
\begin{equation}
	\widehat{Z} = Z,
\end{equation}
show that
\begin{equation}
	[\widehat{Z},\widehat{X}] = -\widehat{Y}
	\hspace{50pt}
	\text{and}
	\hspace{50pt}
	[\widehat{Z},\widehat{Y}] = \widehat{X}
\end{equation}
and
\begin{equation}
	[\widehat{X},\widehat{Y}] = -\widehat{Z}.
\end{equation}\\

Calculate the Hamiltonians:
\begin{equation}
	\alpha(\widehat{Z})
	\hspace{50pt}
	\text{,}
	\hspace{50pt}
	\alpha(\widehat{X})
	\hspace{50pt}
	\text{, and}
	\hspace{50pt}
	\alpha(\widehat{Y}).
\end{equation}

\section{Exercise 2: Other Planimeters}

Watch ``How to Integrate with an Ax?'' by the Mathologer on YouTube.

\subsection{The linear planimeter (exact)}

Derive the canoncial one-form and calculate the kinematic and canonical transformations of the linear planimeter.

\subsection{The hatchet planimeter (approximate)}

Helgoland and The Legend of Fosite.\\

Show that the hatchet planimeter has differential transformations isomorphic to the Euclidean group, $\mathrm{Euc}(2) \cong \R^2 \rtimes \SO(2)$.\\

Show that $\mathrm{Euc}(2)$ is approximately $\mathrm{WH}$ for areas much smaller than the squared length of the hatchet.
This is similar to how the Poincare group, $\mathrm{Poin}(1,1)$, is approximately the Galilean group, $\mathrm{Gal}(1,1)\cong \mathrm{WH}$, in the non-relativistic limit.

\end{document}